\documentclass{aa}

\usepackage{natbib}
\usepackage{graphicx}
\usepackage{graphics}
\usepackage{epsfig,epsf}
\usepackage{amssymb}
\newcommand{\mum}{$\mu$m}
\newcommand{\mums}{$\mu$m\ }
\newcommand{\scanam}{\texttt{Scanamorphos}}
\newcommand\T{\rule{0pt}{3.1ex}}% Adds a space between the text and the [T]op \hline
\newcommand\B{\rule[-1.7ex]{0pt}{0pt}}% Adds a space between the text and the [B]ottom \hline

\begin{document}

\title{The {\it Herschel} \thanks{Herschel is an ESA space observatory with science instruments provided by European-led Principal Investigator consortia and with important participation from NASA.} Exploitation of Local Galaxy Andromeda (HELGA). I: Global far-infrared and sub-mm morphology}

\author{J. Fritz\inst{1}, G. Gentile\inst{1}, M.W.L. Smith\inst{2}, W.K. Gear\inst{2}, R. Braun\inst{3}, J. Roman Duval\inst{4}, G.J. Bendo\inst{5}, M. Baes\inst{1}, S.A. Eales\inst{2}, J. Verstappen\inst{1}, J.A.D.L. Blommaert\inst{6}, M. Boquien\inst{7,8}, A. Boselli\inst{8}, D. Clements\inst{9},  A.R. Cooray\inst{10}, L. Cortese\inst{11}, I. De Looze\inst{1}, G.P. Ford\inst{2}, F. Galliano\inst{12}, H.L. Gomez\inst{2}, K.~D. Gordon\inst{4,1}, V. Lebouteiller\inst{12}, B. O'Halloran\inst{9},  J. Kirk\inst{2}, S.C. Madden\inst{12}, M.J. Page\inst{13}, A. Remy\inst{12}, H. Roussel\inst{14}, L. Spinoglio\inst{15}, D. Thilker\inst{16}, M. Vaccari\inst{17}, C.D. Wilson\inst{18}, C. Waelkens\inst{6}}

\institute{Sterrenkundig Observatorium, Universiteit Gent, Krijgslaan 281, B-9000 Gent, Belgium\\
	\email{jacopo.fritz@ugent.be}
	\and 
         School of Physics and Astronomy, Cardiff University, Queens Buildings, The Parade, Cardiff CF24 3AA, UK.
         \and
         CSIRO Astronomy and Space Science, P.O. Box 76, Epping, NSW 1710, Australia
         \and
         Space Telescope Science Institute, 3700 San Martin Drive, Baltimore, MD 21218
         \and
         UK ALMA Regional Centre Node, Jodrell Bank Centre for Astrophysics, School of Physics and Astronomy, University of Manchester, Oxford Road, Manchester M13 9PL, United Kingdom
	\and 
	Instituut voor Sterrenkunde, K.U.Leuven, Celestijnenlaan 200D, B-3001 Leuven, Belgium
         \and
         University of Massachusetts, Department of Astronomy, LGRT-B 619E, Amherst, MA 01003, USA
         \and
         Laboratoire d'Astrophysique de Marseille, UMR 6110 CNRS, 38 rue F. Joliot-Curie, F-13388 Marseille, France
         \and
	Astrophysics Group, Imperial College, Blackett Laboratory, Prince Consort Road, London SW7 2AZ, United Kingdom
         \and
         Center for Cosmology and the Department of Physics \& Astronomy, University of California, Irvine, CA 92697, USA
         \and
         European Southern Observatory, Karl-Schwarzschild Str. 2, 85748 Garching bei Muenchen, Germany
          \and
         AIM, CEA/Saclay, Orme des Merisiers batiment 709 91191 Gif-sur-Yvette France
         \and
         Mullard Space Science Laboratory, University College London, Holmbury St. Mary, Dorking, Surrey RH5 6NT, UK
         \and
         Institut d'Astrophysique de Paris, UMR7095 CNRS, Universit\'e Pierre \& Marie Curie, 98 bis Boulevard Arago, 75014 Paris, France
	\and
         Istituto di Fisica dello Spazio Interplanetario, INAF, Via Fosso del Cavaliere 100, I-00133 Roma, Italy
         \and
	Dept. of Physics \& Astronomy, Johns Hopkins University, 3400 N. Charles St., Baltimore, MD 21218
	\and
	University of the Western Cape Private Bag X17, Bellville, 7535, Cape Town, South Africa
	\and
	Department of Physics \& Astronomy, McMaster University, Hamilton, Ontario L8S 4M1, Canada
}

\date{Received ...; accepted ...}

\titlerunning{HELGA}
\authorrunning{J. Fritz}

\abstract 
% context 
{We have obtained Herschel images at five wavelengths from 100 to 500 \mums of a $\sim 5.5\times2.5$ degree area centred on the local galaxy M31 (Andromeda), our nearest neighbour spiral galaxy, as part of the Herschel guaranteed time project  ``HELGA''. The main goals of HELGA are to study the characteristics of the extended dust emission, focusing on larger scales than studied in previous observations of Andromeda at an increased spatial resolution, and the obscured star formation.}
%aims
{In this paper we present data reduction and Herschel maps, and provide a description of the far-infrared morphology, comparing it with features seen at other wavelengths.}
%methods
{We use high--resolution maps of the atomic hydrogen, fully covering our fields, to identify dust emission features that can be associated to M31 with confidence, distinguishing them from emission coming from the foreground Galactic cirrus.}
%results
{Thanks to the very large extension of our maps we detect, for the first time at far-infrared wavelengths, three arc-like structures extending out to $\sim 21$, $\sim 26$ and $\sim 31$ kpc respectively, in the south-western part of M31. The presence of these features, hosting $\sim 2.2\times 10^6$ M$_\odot$ of dust, is safely confirmed by their detection in HI maps. Overall, we estimate a total dust mass of $\sim5.8\times 10^7$ M$_\odot$, about 78\% of which is contained in the two main ring-like structures at 10 and 15 kpc, at an average temperature of 16.5 K. We find that the gas-to-dust ratio declines  exponentially as a function of the galacto-centric distance, in agreement with the known metallicity gradient, with values ranging from 66 in the nucleus to $\sim 275$ in the outermost region.}
%conclusion
{Dust in M31 extends significantly beyond its optical radius ($\sim 21$ kpc) and what was previously mapped in the far-infrared. An annular--like segment, located approximately at R$_{25}$, is clearly detected on both sides of the galaxy, and two other similar annular structures are undoubtedly detected on the south-west side even further out.}


\maketitle

%%%%%%%%%%%%%%%%%%%%%%%%%%%%%%%%%%%%%%%%%%%%%%%%%%%%%%%%

\section{Introduction}
An overall view of the interstellar medium of the Milky Way is hampered by the huge column densities that any line of sight has to go through. Andromeda is the closest galaxy offering broadly similar characteristics and an unencumbered view, thus probably being the best target to test our knowledge of the physical processes that govern the formation and evolution of massive spiral galaxies. In fact, high quality data of M31 are available at a wealth of different spectral ranges: X-rays \citep{pietsch05}, ultraviolet \citep{thilker05}, optical (e.g. SDSS, H$\alpha$ by \citealt{devereux94} or the PAndAS survey, see e.g. \citealt{mcconnachie08}), near and far-infrared (FIR) \citep{haas98,barmby06,gordon06,beaton07,krause11}, millimeter \citep{nieten06} and radio \citep[e.g.][]{thilker04,braun09}. Despite this, the observational coverage in the sub-millimeter domain was until now limited by the absence of data longward of  170 \mum, except for the extremely low resolution (with a beam size larger than $40'$) data from the Diffuse Infrared Background Experiment (DIRBE) at 240 \mums \citep{odenwald98}. 

This situation has left M31 unexplored at far-infrared and sub-mm wavelengths in adequate spatial resolution. The advent of the {\it Herschel} Space Observatory \citep{pilbratt10}, with its two instruments PACS (Photodetector Array Camera and Spectrometer) \citep{poglitsch10} and SPIRE (Spectral and Photometric Imaging Receiver) \citep{griffin10}, allows us to fill this observational gap and complete the view of our neighbour galaxy down to sub-mm wavelengths. The other important advantage of {\it Herschel} is the gain in terms of spatial resolution which, at 160 \mum, is about 2.5 times higher than that of the Multiband Imaging Photometer for {\it Spitzer} (MIPS) at similar wavelengths, making it possible to resolve physical scales of about 50 pc at the distance of Andromeda.

On the one hand, M31 is considered in many respects to be very similar to the MW, sharing some characteristics  and similarities such as the Hubble type, the luminosity, and gas content in the respective disks. The fact that it is so close makes it not only the ideal candidate for a comparison with similar studies and ongoing surveys taking place with {\it Herschel} on our own Galaxy \citep[such as Hi-Gal,][]{molinari10}, but also the ideal place for studying the interstellar medium of a spiral galaxy. Hinting at some common aspects between the MW and M31, \cite{yin09} find very similar gas mass fraction profiles (similar scale lengths but different normalizations) and abundance gradients in the two disks.

On the other hand, some differences are evident between the two galaxies: not only does M31 have about twice the baryonic mass of the MW, it has also a larger disk, by a factor of about 2.4. The mass of gas contained in the respective disks, however, is remarkably similar. An overview of the main properties of the two galaxies can be found in \cite{yin09} (see their Table 1).

So far, all the studies of M31's far-infrared Spectral Energy Distribution (SED), mainly relying on data shortwards of $\sim160$ \mum, have speculated about the possible presence of a very cold component ($\sim 10$ K). Such a component would have a SED that will peak at wavelengths longer than 200 \mums \citep{haas98,montalto09}. Due to the wavelength limitations of the data, no constraints could be put on its presence and, even less, its characteristics. The estimates of the dust mass were therefore believed to be only lower limits to the real value. With {\it Herschel} data we are in the position of setting more stringent constraints both to the presence and spatial distribution of a very cold dust component, and its emissivity.

The data we exploit in our survey allow us to also study the dust morphology in the very outer regions of the galaxy. The far-infrared morphology of Andromeda has been the topic of several studies exploiting, over the years, the available infrared facilities such as the Infra Red Astronomical Satellite \citep[IRAS,][]{habing84,xu96}, the Infrared Spatial Observatory \citep[ISO,][]{haas98} and {\it Spitzer} \citep{barmby06,gordon06} each facility improving in both sensitivity and spatial resolution. Evidence for a complete dusty ring at a radius of $\sim 10$ kpc came from the very first  IRAS observations, while following studies exploiting higher resolution were able to detect details such as a discontinuity, most likely caused by an interaction with the satellite galaxy M32 \citep[see, e.g.][]{gordon06}. A fainter ring-like structure at $\sim 15$ kpc was already noted by \cite{haas98} and confirmed by \cite{gordon06}, but nothing more could be said on larger radii because of
  the size limitations of the maps.

While dust located beyond the regions sampled by such observations was indeed detected indirectly by studying the characteristics of the stellar populations in the extreme outskirts of M31 \citep[e.g.][]{cuillandre01,bernard12}, it had never been possible to study its properties and quantify its presence before.

In this work we describe {\it Herschel} PACS and SPIRE observations of M31, at 100, 160, 250, 350 and 500 \mum, extending out to about 34 kpc ($\sim2.5^\circ$) from its centre, presenting the most complete far-infrared/submillimeter survey of Andromeda. We describe the observations and the data reduction steps in \S \ \ref{sec:datared}, while in \S \ \ref{sec:ancillary} we give a general overview of the ancillary dataset that we exploit in our project, together with a comparison to {\it Herschel} maps. In \S \ \ref{sec:results} we use high and low-resolution {\sc Hi} maps, entirely covering our fields, to disentangle the dust dynamically connected to Andromeda from the foreground Galactic cirrus, and in \S \ \ref{sec:dust} we use these foreground--corrected maps to study the dust properties. Conclusions and remarks are given in \S \ \ref{sec:conclusions}.

Throughout this paper and in all works within the HELGA project, we will assume a distance to Andromeda of $785$ kpc \citep{mcconnachie05}, which turns into a linear scale of 3.8 pc/$''$. As for the inclination and position angles, we are using $i=77^\circ$ and PA$=38^\circ$, respectively.

\begin{figure*}
\centering
\includegraphics{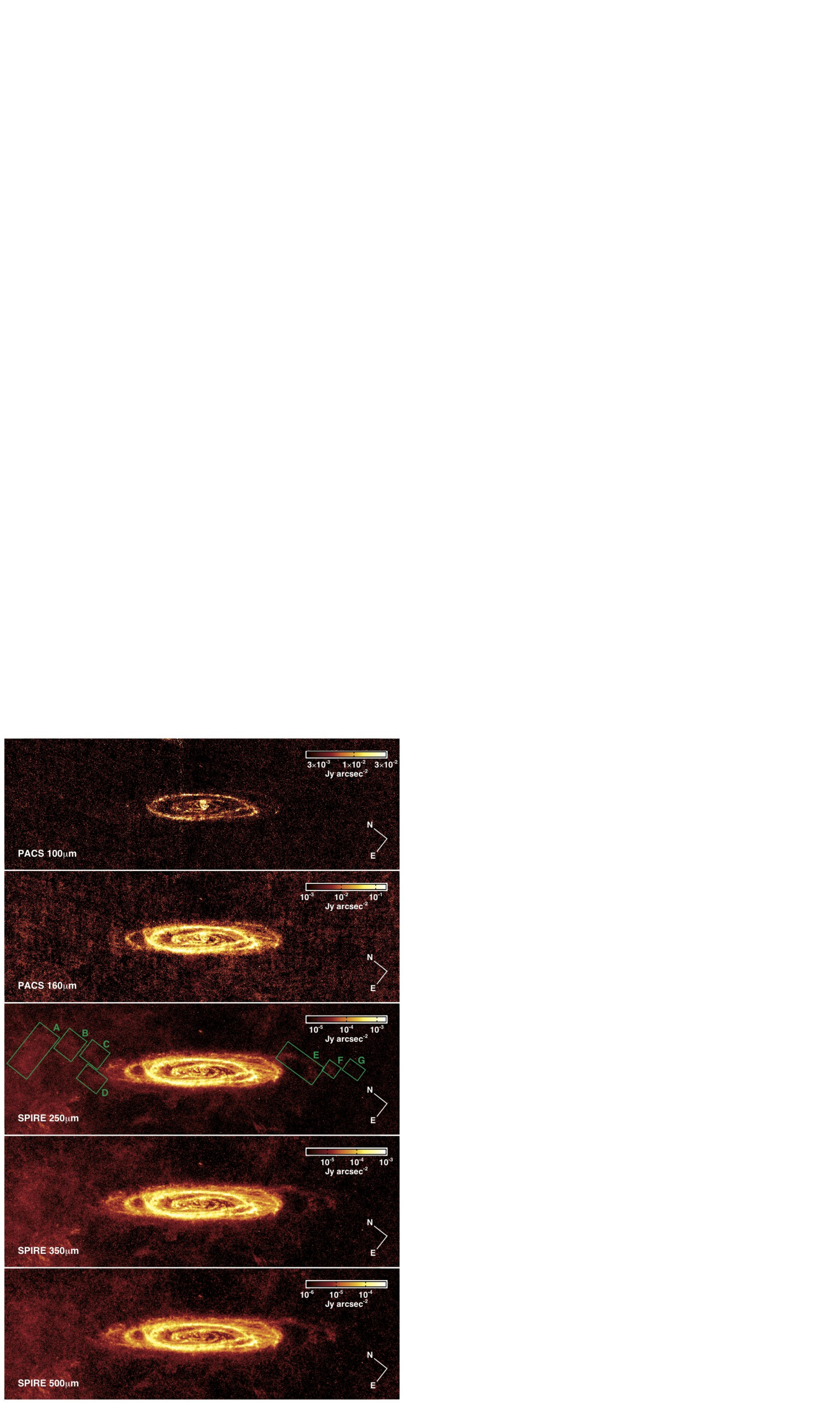}
\caption{Images in the five {\it Herschel} bands (from top to bottom: PACS 100 and 160 \mum, SPIRE 250, 350 and 500 \mum). These are the original images, where no correction for the foreground cirrus was applied (see Sec. \ref{sec:identification}). The green rectangles in the 250 \mums map represent tentative detections of dust in the outskirts (see discussion in Sec. \ref{sec:fir_struct}). The directional axes length corresponds to $15'$ ($\sim 3.42$ kpc).}
\label{fig:herschelimages}
\end{figure*}
 
\section{Observations and data reduction}\label{sec:datared}
In this section we describe the observing strategy used to obtain the maps presented in this paper, and the data reduction techniques. Furthermore, given the greater difficulty of removing instrumental artefacts in PACS data, some quality checks are performed for the latter.

\subsection{Observations strategy}
We observed a field of $\sim 5.5^\circ \times 2.5^\circ$ centered on M31, with both PACS and SPIRE in parallel mode, at a scanning speed of $60''/$s. PACS data were taken at 100 and 160 \mum. Images at the five wavelengths are presented in Fig. \ref{fig:herschelimages}. To obtain such a large map, we split our observations in 2 parts, A and B, each  observed with a cross-scan, so that the final map is a composition of four observations. Each one covers a field about $3^\circ \times 2.5^\circ$ wide, displaced approximatively along the major axis of the galaxy, and they overlap at the nucleus. The fields are centered at RA=0h37m38.870s, DEC=+40d00m51.30s (A\footnote{OBSIDs 1342211294 and 1342211309}) and RA=0h48m01.670s DEC=+42d30m33.20s (B\footnote{OBSIDs 1342211319 and 1342213207. Data are available at: \texttt{http://herschel.esac.esa.int/Science\_Archive.shtml}}), and they both observe a stripe, parallel to the galaxy's minor axis, which is about $0.2^\circ$ and $\sim0.3^\circ$ wide in PACS and SPIRE maps, respectively. These observations were scheduled on the 18th, 19th, 20th and 21st of December, 2010. Due to some issues with the star trackers, the last orthogonal scan of the northern portion of the map, was lost. The field was hence re-observed on the 23rd of January, 2011, so that it displays a slight rotation with respect to the nominal scan. 

\subsection{PACS data reduction and quality check}\label{sec:pacs}
\indent The basic PACS data reduction (up to level-1 where pixel flagging, flux density conversion, and sky coordinate association for each pixel of the detector are applied) was performed in the {\it  Herschel} Interactive Processing Environment (HIPE) {\bf v8}, which includes the most recent responsivity factors. At this stage, the PACS timelines are still dominated by brightness drifts caused by low-frequency noise. In order to subtract these drifts, remove glitches, and project the timelines on the final map, we applied the IDL algorithm \scanam \ \citep[version {\bf 15},][]{roussel11} to the level-1 PACS timelines. To this end, it exploits the fact that each sky pixel is observed multiple times by several different detectors, avoiding assumptions or modeling of the noise properties. The final PACS maps have a pixel scale of $2''$ and $3''$ in the 100 (green) and 160 \mums (red) band maps, respectively, and the FWHM of point sources is about $12.5''$ and $13.3''$ in the green and red channel, respectively, (see the ``PACS photometer point spread function'' report, by D. Lutz\footnote{November 3, 2010, version 1.01: \texttt{herschel.esac.esa.int/} \texttt{twiki/pub/Public/PacsCalibrationWeb/bolopsfv1.01.pdf}}). Note that, for high scan speed observations, the point spread function (PSF) is elongated in the scanning direction. This yields a spatial resolution on a physical scale of about 48.1 pc at 100 \mums and 50.7 pc at 160 \mum..

In order to examine the accuracy of the PACS calibration and of the flux output from \scanam, we performed a pixel-to-pixel comparison between the PACS 160 \mums and the {\it Spitzer}  MIPS 160 \mums images on the one hand, and between the PACS 100 \mums and DIRBE \citep{hauser98} at the same wavelength. We first convolved the PACS 160 \mums map (resp. PACS 100 \mums map) to the MIPS 160 \mums (resp. DIRBE 100 \mum) resolution using the convolution kernel provided by \cite{aniano11} (a gaussian kernel was used to smooth 100 \mums data to the DIRBE resolution), and then resampled the maps on the same astrometric grid using the \texttt{hastrom}  task in IDL. Finally, we compared the flux in each pixel of the map. PACS 100 \mums and DIRBE 100 agree within 10\%, as found in \cite{meixner12}. As for the 160 \mums band, we confirm the non--linear behavior of MIPS above 50 MJy/sr, which was also found by \cite{stansberry07} and is discussed in detail in \cite{meixner12}. Below this threshold, PACS and MIPS agree within 5\%, whereas the ratio PACS/MIPS is non-linear above that, leading to differences in brightness up to 30\%, with MIPS underestimating the fluxes. Overall, the PACS extended calibration can be considered to be accurate to within 10\%. A comparison of the total flux of M31 measured from PACS and MIPS maps at 160 \mum, indeed yields an agreement to within about 5\%. We have also investigated different baseline subtraction methods \citep[HERITAGE-like, see][]{meixner12} and found similar results. 

Hence, in the rest of this work, we will adopt a 10\% value as uncertainty in PACS fluxes at both wavelengths.
 
\subsection{SPIRE data reduction}\label{sec:spire}
The SPIRE data were processed up to Level-1 (i.e. to the level where the pointed photometer time-lines have been derived) with a custom driven pipeline script adapted from the official pipeline ({\sl POF5\_pipeline.py}, dated 8 Jun 2010) as provided by the SPIRE Instrument Control Centre (ICC)\footnote{See ``The SPIRE Analogue Signal Chain and Photometer Detector Data Processing Pipeline'' \cite{griffin09} or \cite{dowell10} for a more detailed description of the pipeline and a list of the individual modules.}.

This Jython script was run in HIPE coming with the continuous integration build number 4.0.1367. In terms of the SPIRE scan map pipeline up to Level-1, this was very similar to the {\it Herschel} Common Science System/Standard Product Generation v5 since we already had patched the new parts of v5 into our old v4. So we used the new flux calibration product of v5 where the SPIRE calibration is finally based on Neptune data, as well as the {\tt concurrentGlitchDeglitcher}: the latter removes the effect of often low level, but frequent glitches appearing in all bolometers of an array at the same time. 

The differences from the standard pipeline are that we used the \texttt{sigmaKappaDeglitcher} task instead of the ICC-default \texttt{waveletDeglitcher} and we did not run the default \texttt{temperatureDriftCorrection} and the residual, median baseline subtraction. Instead we use a custom method called BriGAdE \citep{smith13}, to remove the temperature drift and bring all bolometers to the same level (equivalent to baseline removal). BriGAdE removes the temperature drift by fitting the thermistor timeline to the bolometer signal timeline assuming a linear relationship to find the baseline to subtract \citep[a similar process is presented in][]{pascale10}. In addition to BriGAdE's standard source masking we also manually mask Andromeda from the timeline fitting and adjust the length of the linear fit to every second scan-leg (including turn-around) to avoid issues interpolating the thermistor fitting process for the two observations where the satellite turns-around while pointing at the centre of Andromeda. For the two northern observations a gradient in the cirrus was found (as seen in previous IRAS observations) in the north-south direction, however due to baseline subtraction this is not seen in the east-west observation. To compensate for this we use the IRAS map to identify a roughly constant level of cirrus across the two observations and use these regions to perform our thermistor fitting. The northern and southern observations are then normalised to have the same zero level and mosaicked to create the final map. We have found this method improves the baseline subtraction significantly, especially in cases where there are strong temperature variations during the observation.
\begin{figure*}
\centering
\includegraphics{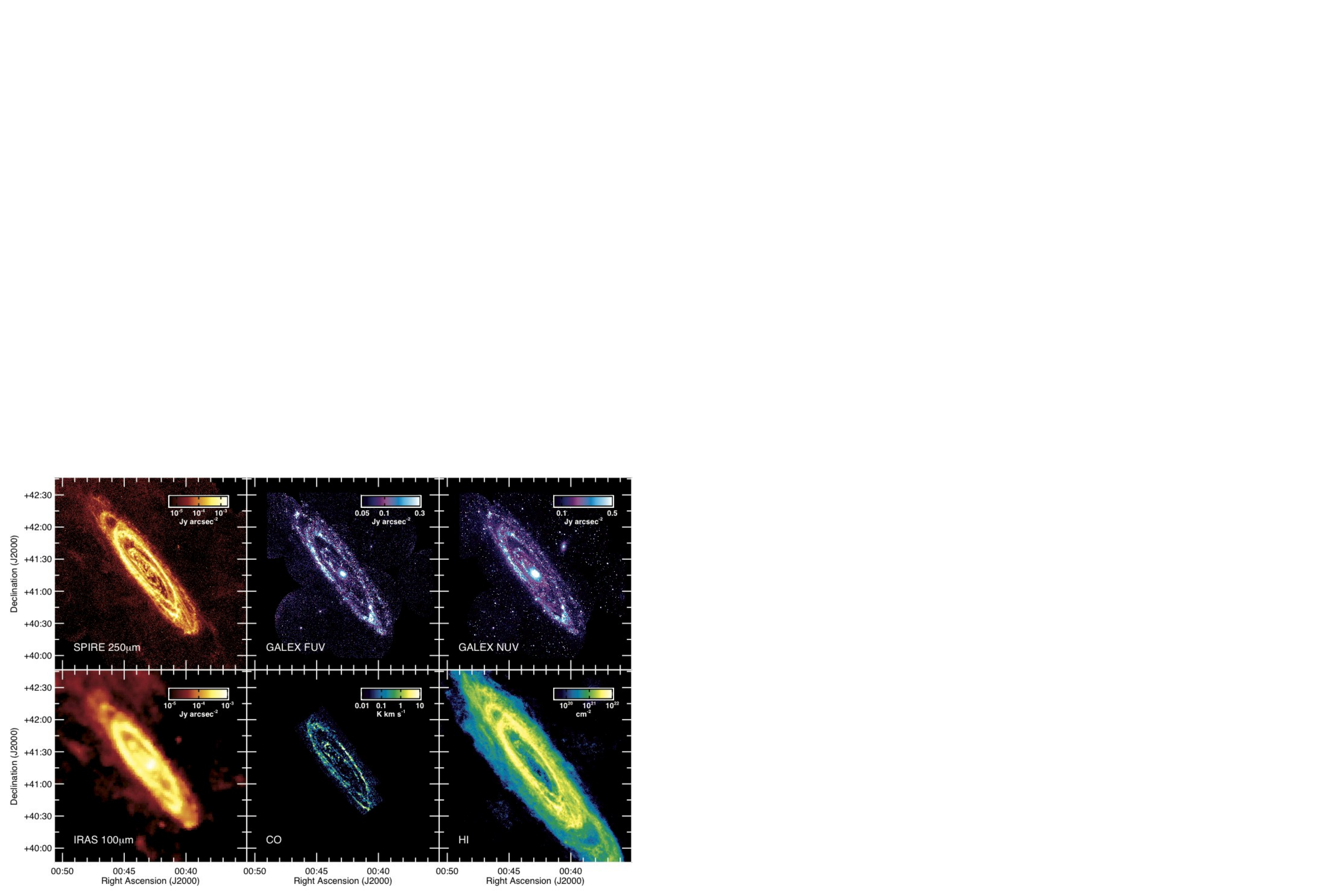}
\caption{Upper row, from left: the SPIRE 250 \mums image (this work), and far and near--ultraviolet images taken with GALEX \citep[][]{thilker05,gildepaz07}. Lower row, from left, the {\it IRAS} 100 \mums (MJy/sr) (shown as a comparison), the CO map \cite[by][in K km/s]{nieten06}, and the {\sc Hi} column density map (expressed in cm$^{-2}$), from \cite{braun09}. All the images are set to the same spatial scale.}
\label{fig:ancillary}
\end{figure*}

Our final maps were created using the naive mapper provided in the standard pipeline using pixel sizes of $6''$, $8''$, and $12''$ at 250, 350, and 500 \mum, respectively. The non--standard pixel sizes were chosen in order for the pixels to have approximatively the size of $1/3$ of the beam. The FWHM of the SPIRE beams vary as a function of the pixel size and are $18.2''$, $24.5''$, and $36.0''$ at 250, 350, and 500 \mum, for the pixel scale we have used \citep{swinyard10}. In addition the 350 \mums measurements are multiplied by 1.0067 to update our flux densities to the latest v7 calibration product. The calibration uncertainties are taken to be 5\% correlated error between bands with a 2\% random uncertainty, however it is recommended that these values are added linearly instead of in quadrature (see also SPIRE user manual\footnote{\texttt{ http:// herschel.esac.esa.int/Docs/SPIRE/pdf/ spire\_om.pdf}}).

\section{Ancillary data}\label{sec:ancillary}
In this section we first give a brief overview of the multi--wavelength dataset that is already available in the literature and that will be exploited by our project in future papers. We then  present a comparison of {\it Herschel} data with other wide-field observations at various wavelengths, focusing in particular on the {\sc Hi} datasets.

In Fig. \ref{fig:ancillary} we show maps at UV, FIR (IRAS at 100 \mum), millimeter ($^{12}$CO(1-0)) and radio ({\sc Hi}) wavelengths, compared to SPIRE 250 \mums (upper--left panel). GALEX images \citep{thilker05,gildepaz07}, with a PSF ranging from $4''$, for Far-UV band, to $5''$ for the Near-UV (FWHM), span $\sim 3^\circ$ along the major axis, and they also include the satellite galaxy NGC205. In both maps the 10 kpc {\bf ($\sim 0.75^\circ$)} ring is clearly traced by UV emission, which extends out to the 15 kpc {\bf ($\sim 1.1^\circ$)} ringlike structure. The major difference from a morphological point of view, is found in correspondence of the ``hole'' which splits the ring \citep{gordon06,block06}, which in UV images is peaking towards the outskirts. Otherwise UV and IR emission are well mixed across the galaxy.

M31 was observed by {\it Spitzer} with both the InfraRed Array Camera (IRAC) \citep[][with a FWHM of $1.9''$]{barmby06}, and with MIPS \citep[][PSFs from $6''$ to $40''$]{gordon06}. The area imaged by the {\it Spitzer} survey was a $\sim 3\times 1^\circ$ region, about the same size as that of the UV data.

CO data \citep{nieten06}, used to trace the molecular gas phase, have a spatial resolution of $23''$ and only cover a $\sim 1.8^\circ \times 0.5^\circ$ wide region, sampling the whole 10 kpc ring. As expected, CO emission shows a remarkable morphological similarity with IR structures: faint nuclear emission with most of the radiation being emitted by the 10 kpc ring. 

Finally, {\sc Hi} maps, crucial to distinguish whether the dust belongs to M31 or to the MW, are discussed in detail in Section \ref{sec:HIdata}: both sets \citep{thilker04,braun09} fully include the whole HELGA field (see e.g. Fig. \ref{fig:ident_paper}).

\subsection{The {\sc Hi} dataset}\label{sec:HIdata}
Exploiting the velocity information carried by {\sc Hi} observations  is a commonly used means to disentangle FIR emission associated with Galactic cirrus, from that coming from extragalactic sources. Given the relatively low galactic latitude of Andromeda ($b\simeq-22^\circ$) contamination by cirrus emission may be important in the field and has to be identified. Two datasets are used for this purpose: a high-resolution cube, presented in \cite{braun09}, and a low-resolution cube, presented in \cite{thilker04}. Whenever possible, we used the high-resolution one. 

The high-resolution cube is derived in \cite{braun09} from mosaic Westerbork Synthesis Radio Telescope (WSRT) observations. The maximum angular resolution is $18'' \times 15''$, which corresponds to a spatial resolution of $\sim 68 \times 57$ pc. The velocity coverage of the created maps ranges from -621 km s$^{-1}$ to -20 km s$^{-1}$, and \cite{braun09} report Galactic emission from -130 km s$^{-1}$ to +45 km s$^{-1}$. However, we also make use of the low-resolution cube obtained by \cite{thilker04} from the Green Bank Telescope (GBT) observations, because (1) their velocity range (from -828 km s$^{-1}$ to +227 km s$^{-1}$) completely covers the range of the foreground Galactic emission, and (2) foreground Galactic emission is expected to be angularly very extended, and GBT data are more sensitive to extended emission than WSRT data. The GBT beam ($9.1'$) corresponds to $\sim 2.1$ kpc at M31 distance.

\section{Results}\label{sec:results}
\subsection{{\bf Foreground emission}}\label{sec:identification}
An attempt to detect diffuse dust structures in the outskirt of nearby galaxies has been previously made by \cite{davies10}, who carried out a detailed analysis of the infrared emission in the M81 group. They looked for an association of the filamentary emission, detected at SPIRE wavelengths in the proximity of M81, with similar structures detected in the optical. To identify the origin of this diffuse emission, they exploited the association of dust with atomic hydrogen and the kinematic information contained in the {\sc Hi} dataset. They found that most of the diffuse far-infrared and optical emission located outside the clear extent of the disk of M81 could be identified with Galactic cirrus. 

We have followed a similar approach: in order to disentangle pristine M31 emission from foreground Galactic (cirrus) emission, we have searched for identifications of the various IR structures visible in Fig. \ref{fig:ident_paper}, with similar structures visible in the high--resolution {\sc Hi} dataset. The latter  has the advantage of carrying kinematic information: in principle, IR structures identified with {\sc Hi} emission at non-Galactic velocities can be safely attributed to M31. However, the situation for M31 is somewhat unfortunate: its systemic velocity is approximately between -310 and -300 km s$^{-1}$ and its maximal rotation velocity is of the order of 250-300 km s$^{-1}$ (the inner rotation velocity, where the peak rotation velocity is situated, is quite uncertain, see e.g. \citealt{chemin09} and \citealt{corbelli10}). This implies that for the receding (north-east) side of M31, where the velocity is close to 0 km s$^{-1}$, disentangling Galactic and M31 emission is potentially problematic.  

\begin{figure*}[!t]
\begin{tabular}{l}
\includegraphics[width=1.0\textwidth]{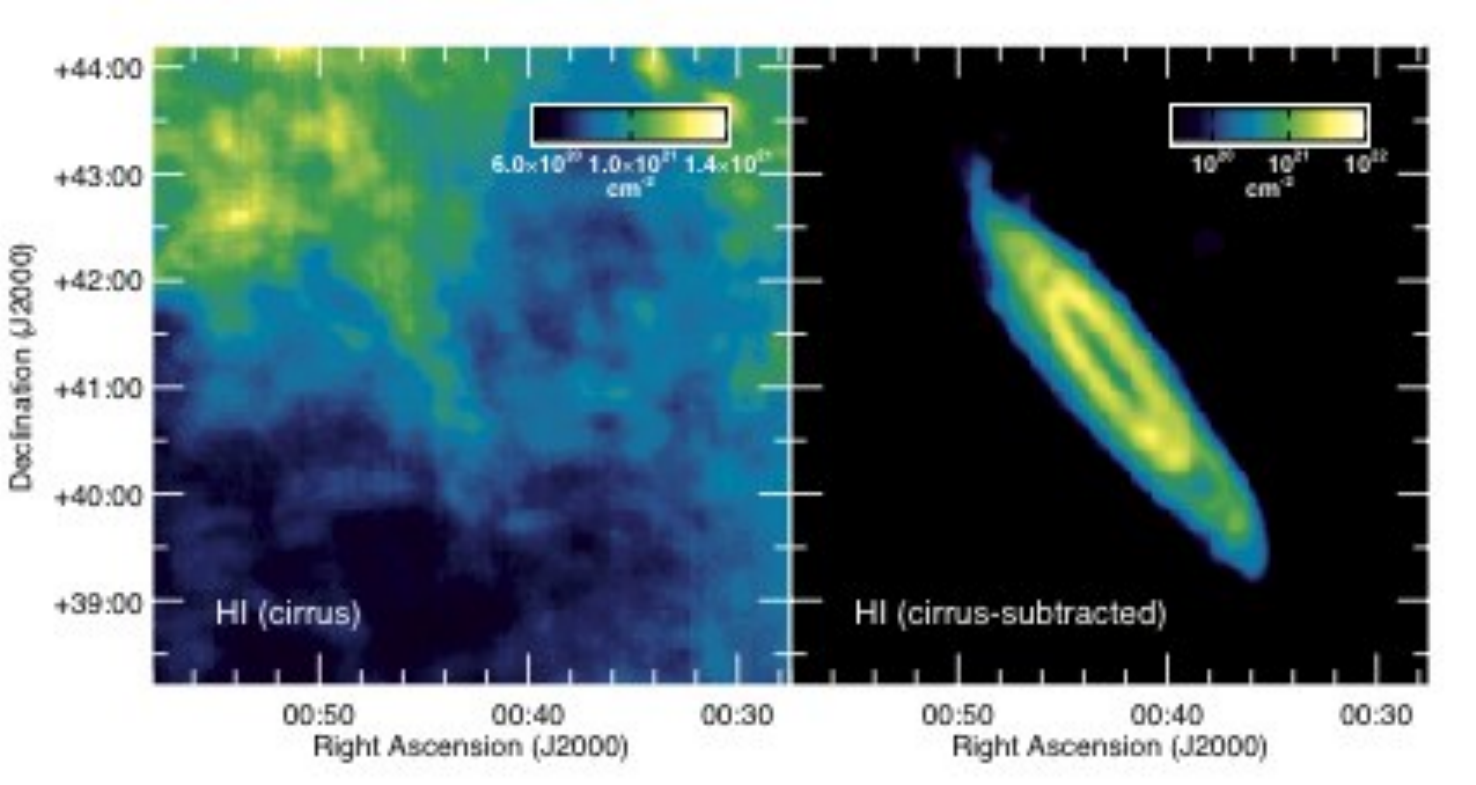} \\
\includegraphics[width=1.0\textwidth]{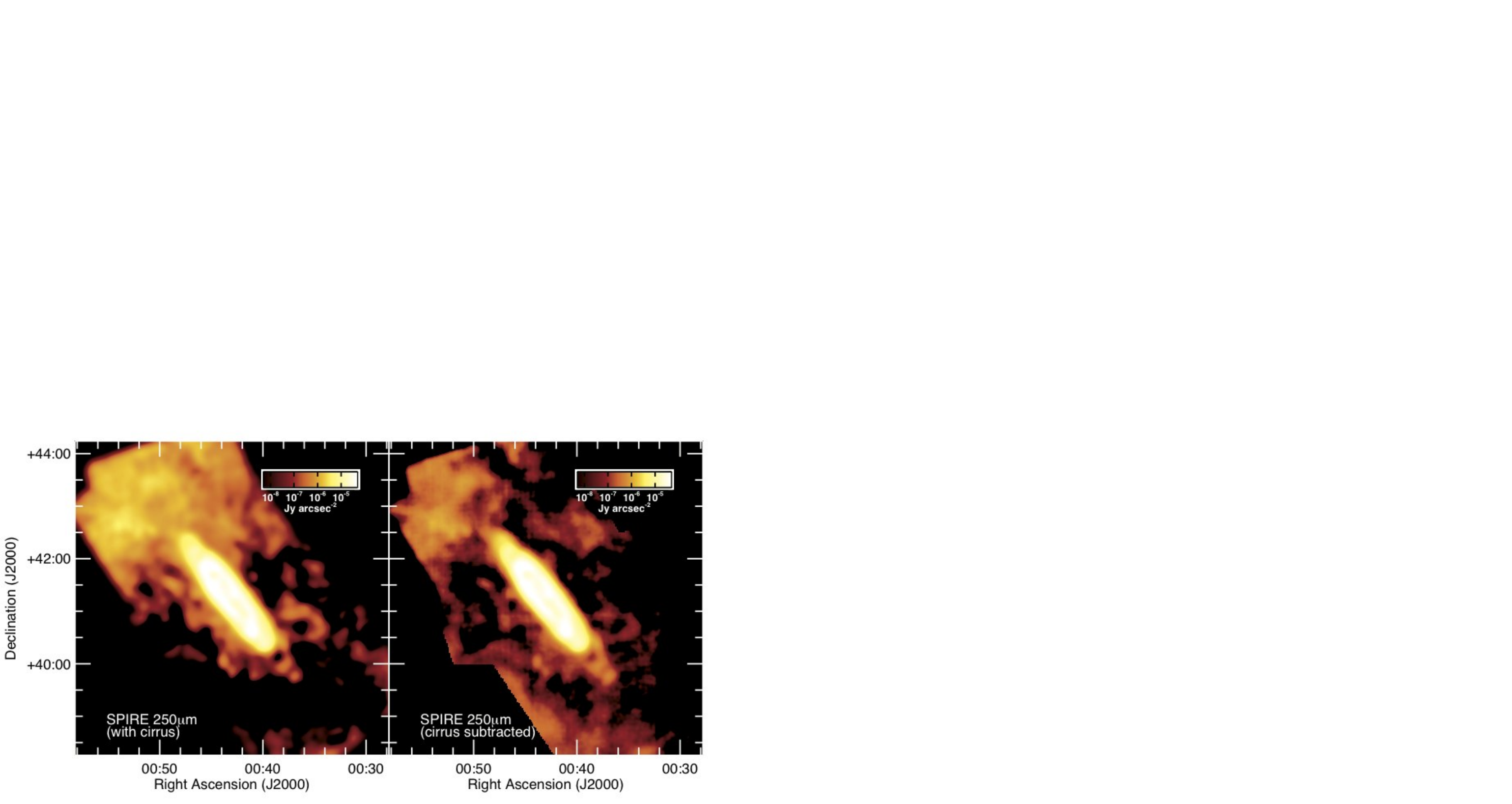}\\
\end{tabular}
\caption{Top row: on the {\it left--hand} panel the {\sc Hi} galactic emission, as derived from GBT observations, is shown (see text for details). On the {\it right--hand} panel, M31 associated {\sc Hi} emission, obtained as the difference between the GBT map integrated over all the velocities, and the pure MW cirrus emission. Bottom row: on the {\it left--hand} panel the original SPIRE 250 \mums image, convolved to the GBT resolution and matched to the same pixelsize, compared to the cirrus subtracted image, at the same resolution ({\it right--hand} side).}
\label{fig:GBT}
\end{figure*}
To try to overcome this issue, we have derived maps of both M31 and the foreground {\sc Hi} emission from the low resolution GBT observations. 

The foreground emission map, shown in the left panel of Fig. \ref{fig:GBT}, was made in the appropriate range of velocities, from $-136$ to $+42$ km s$^{-1}$, that includes the Galactic emission, under the assumption of negligible self-opacity over this interval. This foreground emission map has had regions of M31 emission replaced within each relevant velocity channel, with a polygon containing a constant value related to the median value of the surrounding region at that same velocity. The polygons were defined, as described in \cite{braun09}, to contain all of the contiguously connected {\sc Hi} emission in each velocity channel that could be attributed with confidence to M31. 

The foreground {\sc Hi} image can then be used to determine appropriate scale factors to allow a correction of each of the FIR images. To achieve this, we have explored two different approaches. Firstly, using the {\sc Hi} cirrus map, we have derived maps of the Galactic cirrus emission at {\it Herschel} wavelengths, by assuming a Galactic gas--to--dust ratio, a given dust temperature (we have tried a set of temperatures ranging from 10 to 15 K), and that the dust is emitting as a modified black body with an emissivity value taken from \cite{draine03}. Subtracting these images from SPIRE data to obtain cirrus--free maps, gave quite poor results, yielding heavy over--subtraction, with large portions of the maps displaying negative fluxes after this correction, mostly in the north--east region, the most confused area of the field. 

\begin{figure*}
\begin{tabular}{lll}
\includegraphics[width=0.315\textwidth]{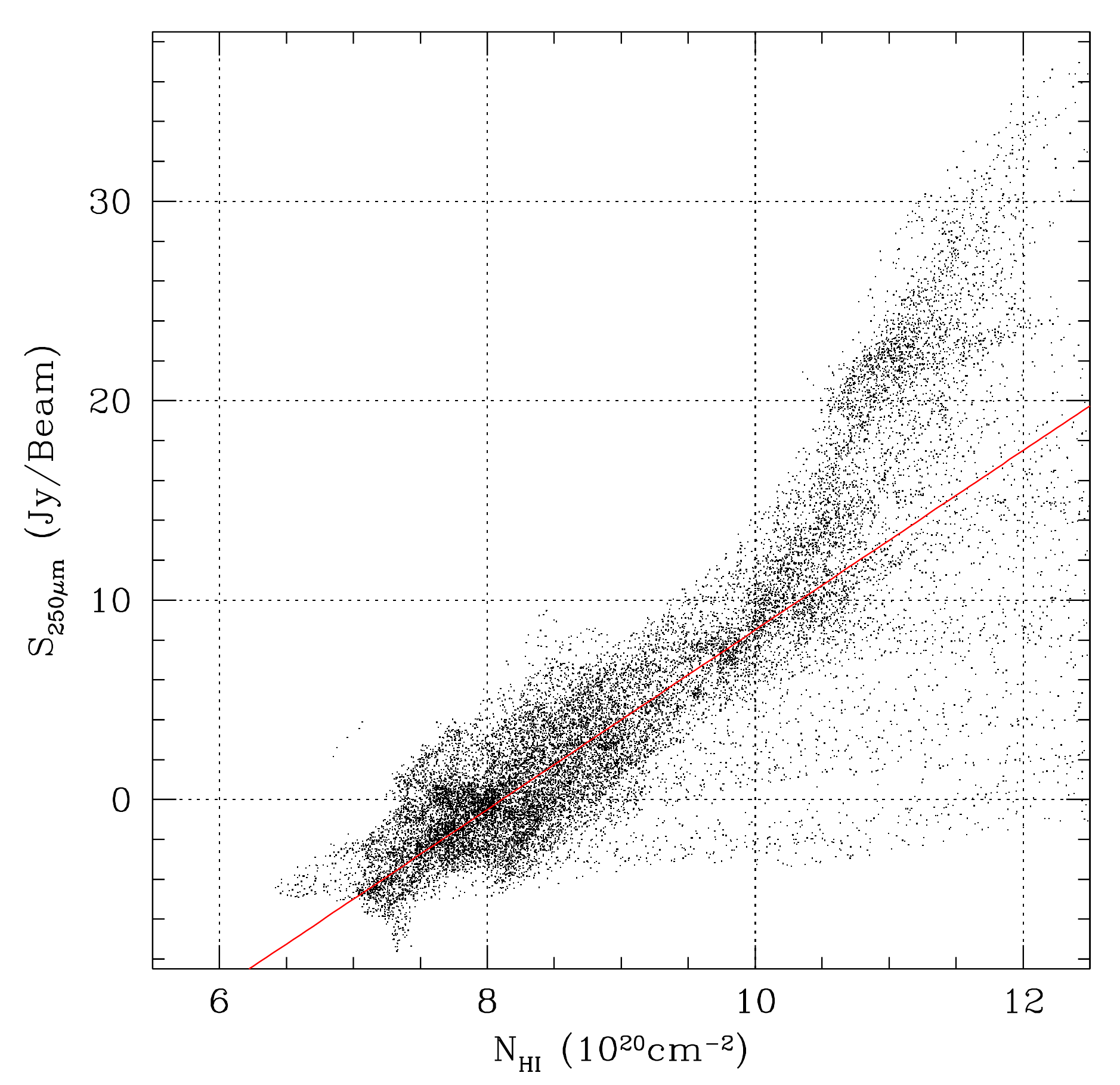} &
\includegraphics[width=0.315\textwidth]{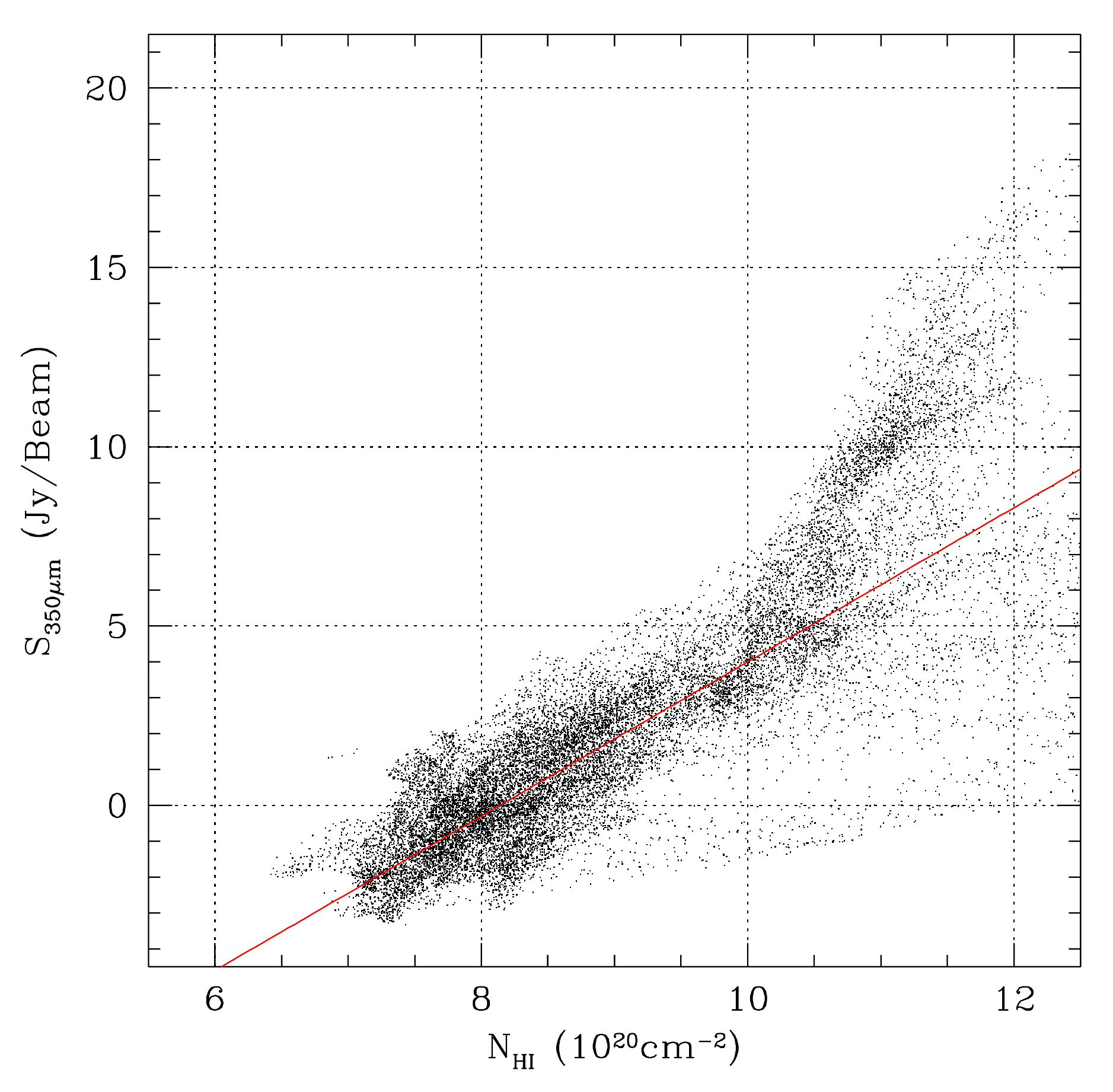} &
\includegraphics[width=0.315\textwidth]{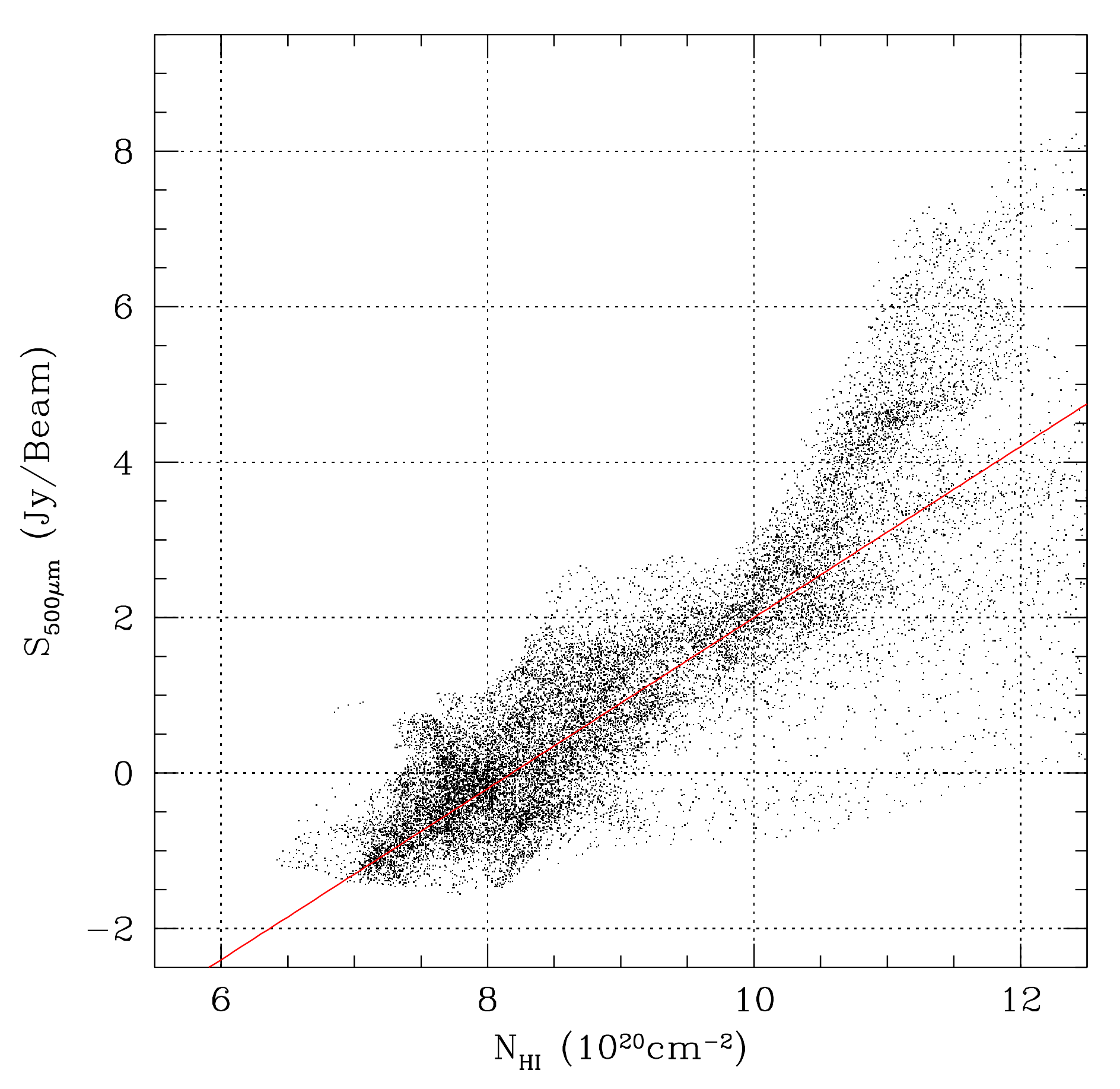}\\
\end{tabular}
\caption{Pixel-by-pixel correlation between {\sc Hi} column density, derived from the full {\sc Hi} integral M31 image at low resolution,  and the flux at 250, 350 and 500 \mums from left to right, respectively. The red lines represent a linear fit to the data.}
\label{fig:scatterFIR}
\end{figure*}

Given the potentially very large parameters space which is involved in this procedure ($\beta$, dust temperature, gas--to--dust ratio, dust emissivity), making the solution to this problem highly degenerate, we have decided to try a different, more empirical, approach. We have looked for a pixel-by-pixel correlation between the FIR emission at SPIRE wavelengths and the {\sc Hi} cirrus emission (Fig. \ref{fig:GBT}, top panels) outside of M31 which we have derived as explained above, in maps that were convolved to a common angular resolution and matched in pixel size. Fig. \ref{fig:scatterFIR} shows the scatter plots of the FIR emission versus {\sc Hi} column density, together with a linear fit to the points, at 250, 350 and 500 $\mu$m, respectively. 

In these plots we note a group of data points displaying a {\sc Hi}--FIR correlation with a clearly different slope with respect to our best fit, above {\sc Hi} column densities of about $10^{21}$ cm$^{-2}$. Most of these points correspond to regions that are clearly associated with MW cirrus emission in the N-E side of Fig. \ref{fig:herschelimages} that is still be visible in the maps even after the correction (e.g. see Fig. \ref{fig:GBT}, right side of the bottom row).

The linear fit was performed in a limited range of N$_{\rm HI}$ and flux values, that was relevant for the foreground estimation. We have experimented extensively with other types of fitting, which included more of the FIR-bright regions, but these always resulted in over-subtraction of the foreground in the most Northern portion of the FIR images. Cirrus maps at the SPIRE wavelengths are hence derived by multiplying each pixel of the {\sc Hi} foreground map by a value derived from the aforementioned correlations. A residual emission in all the FIR maps is still clearly visible in this part of the field (see e.g. Fig. \ref{fig:GBT} and Fig. \ref{fig:ident_paper}), having probably very different dust/gas properties relative to the average Galactic foreground toward the M31 disk. Furthermore, we stress that given the low spatial resolution of the {\sc Hi} map, any cirrus structure smaller than $\sim 9'$ will not be removed. 

The bottom panels of Fig. \ref{fig:GBT} show the comparison between the original 250 \mums map (left panel) and the cirrus-subtracted one (right panel; both images are at the GBT resolution) that has been obtained by applying the pixel-by-pixel correlations mentioned above. The south-east edge of the latter image displays now some additional emission which was not present before the subtraction. This is an artifact, caused by the foreground subtraction, which only affects this particular region. In fact, this region has the lowest {\sc Hi} column density (see upper-left panel of Fig. \ref{fig:GBT}) which, in the $S_{250\mu m}$ vs {\sc Hi} relation (Fig. \ref{fig:scatterFIR}), lies in the negative side of the ordinate axis  (this is due to slight imperfections in the baseline subtraction). Hence, subtracting these negative values gives a positive contribution on this side of the map, but it is not affecting the emission from M31.

The lines that produced the best overall corrections are overplotted in Fig. \ref{fig:scatterFIR}.  The slopes of the lines for the 250, 350 and 500 \mums bands (in data where the beams have been matched to the {\sc Hi} beam) are 4.50, 2.15 and 1.10 Jy/beam/$10^{20}$ cm$^{-2}$, respectively. The relative FIR brightnesses of the foreground in these bands are $1: 0.48 : 0.24$. For comparison, a modified blackbody with a dust temperature of 15 K and and $\beta$ of 2 would give relative brightnesses of $1 : 0.56 : 0.21$, which is reasonably close to the measured values and provides a physical justification for our method. We have used these correlations to derive, from the GBT {\sc Hi} foreground image, FIR maps of the Galactic cirrus emission at all SPIRE wavelengths, that we have then subtracted from the respective {\it Herschel} maps to correct for the presence of foreground dust. This correction only marginally affects M31 emission, yielding a pixel--to--pixel median difference of about 3\% at all SPIRE bands, within a galacto-centric distance of $\sim 20$ kpc. A similar investigation on PACS data yielded no positive results, probably due to the lower sensitivity at these wavelengths. PACS maps are, hence, not corrected for the cirrus emission.

In the rest of this work we use the foreground--corrected maps, unless stated otherwise.

\subsection{Far-Infrared structures}\label{sec:fir_struct}
\begin{figure}
\begin{center}
\includegraphics[width=0.5\textwidth]{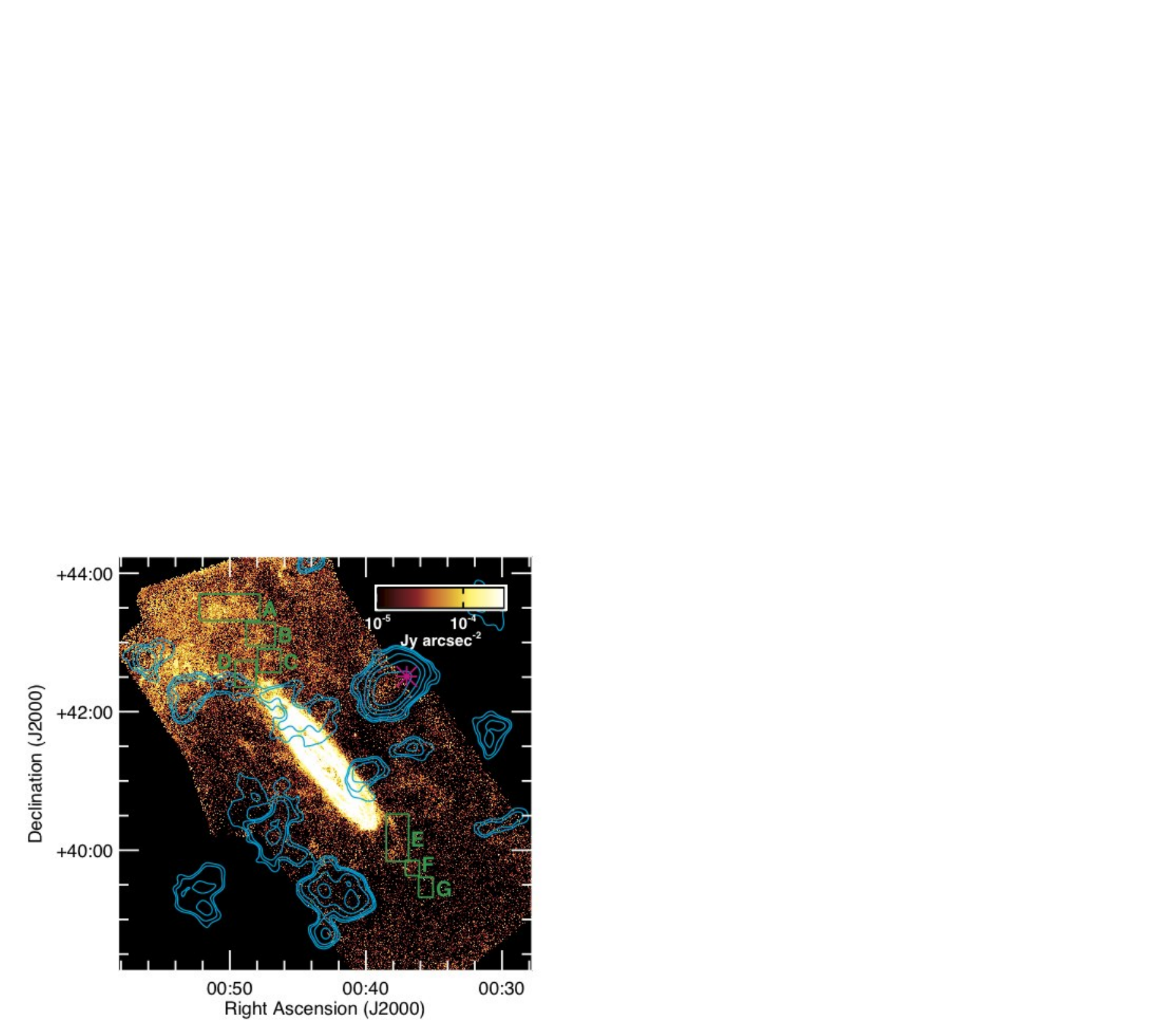}
\end{center}
\caption{The {\it Herschel} 250 \mums image processed for subtracting the foreground cirrus emission (see text), is shown. Dark green rectangles represent identifications of structures of which we investigate the association to M31. Blue contours represent the total column density for discrete and diffuse M31 high-velocity {\sc Hi} clouds, presented in \cite{thilker04} ({\sc Hi} emission from the disk of Andromeda was removed), where we also look for the presence of dust associated to M31. Contours are 0.5, 1, 2, 5, 10, and 20 $\times$ 10$^{18}$ atom cm$^{-2}$. The magenta star identifies Davies' Cloud.}
\label{fig:ident_paper}
\end{figure}

Figure \ref{fig:herschelimages} presents images in the five {\it Herschel} bands. The morphology is characterized by a set of rings and arc-like structures whose prominence varies as a function of the wavelength. The well--known 10 kpc ring \citep{habing84,haas98,gordon06} is the most prominent structure at all wavelengths, while the innermost regions become progressively less luminous at increasing wavelengths. Hints of a ring--shaped structure extending further out ($\sim 15 $ kpc) already seen both in ISO and {\it Spitzer} maps, can be noted at 100 \mum, and its evidence becomes clear in maps at all other wavelengths. Other even more external structures can only be seen in SPIRE images. Two other ring--like structures extending to the south--west are well detected at these frequencies, and they will be discussed below. 

Images at all {\it Herschel} wavelengths are dominated by the 10 kpc ring which is clearly visible up to 500 \mum. Furthermore, extending out to $\sim 15$ kpc from the centre, both PACS and SPIRE maps reveal a ring-shaped structure, whose presence, at infrared wavelengths, had already been noticed by \cite{haas98} and \cite{gordon06} (see their Figs. 1). In the analysis performed within the latter work, which exploits {\it Spitzer} images, it was already questioned whether the far-infrared emission from Andromeda could extend to larger radii than those sampled by their $3^\circ$ map.

\begin{figure*}
\begin{center}
\includegraphics[width=0.95\textwidth]{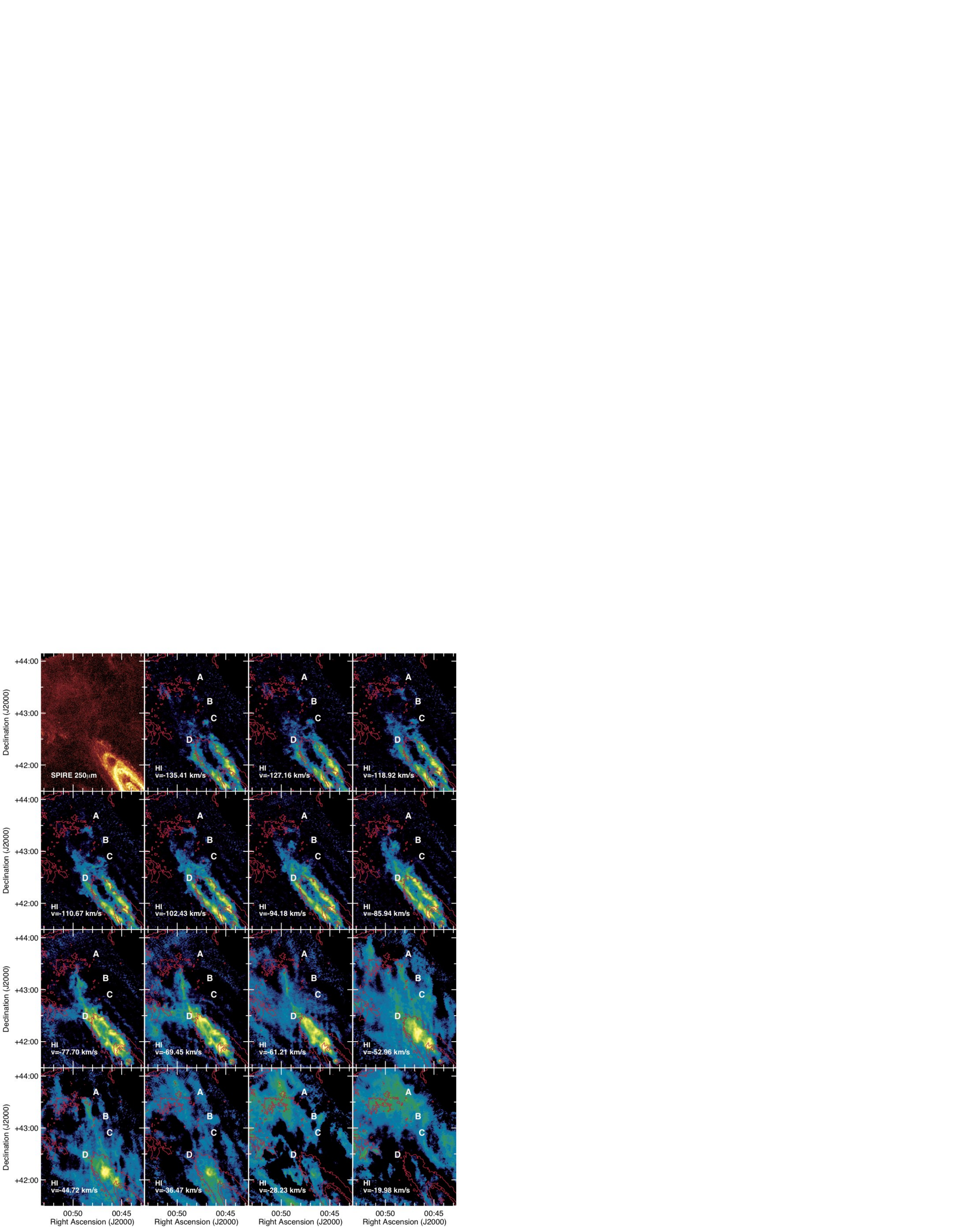}
\end{center}
\caption{Zoom of the north-east edge of M31. The top left panel shows part of the {\it Herschel}  250 \mums image as a colour map, while the same image is shown as contour in the other panels (contours are 3 mJy beam$^{-1}$, i.e. $\sim$ 3.8$\sigma$, 30 mJy beam$^{-1}$, and 300 mJy beam$^{-1}$), superimposed to the {\sc Hi} map at various velocities. The channel velocities are indicated in the bottom left corner of each panel.  Structures in the IR identified as A to D (see also Fig. \ref{fig:herschelimages} and \ref{fig:ident_paper}) are indicated in the panels.  The other panels show the SPIRE 250 \mums image as contours and channel maps from the high-resolution {\sc Hi} datacube as colour map.} 
\label{fig:nnne_all_jpeg}
\end{figure*}

In Fig. \ref{fig:ident_paper} (but see also Fig. \ref{fig:herschelimages}) possibly positive gas--dust associations with the disk and extended components are shown as rectangles. The north-east edge of M31 is contaminated the most by foreground Galactic emission, and hence difficult to analyze, even after correcting for the foreground dust component. The two structures that we have indicated as {\bf C} and {\bf D} in Fig. \ref{fig:ident_paper} are spatially coincident with a structure that is probably associated with M31 and a structure that is probably foreground Galactic emission, respectively. This is illustrated in Fig. \ref{fig:nnne_all_jpeg}. For structure C, this is visible in channels $-135.41$ to $-102.43$ km s$^{-1}$ ({\sc Hi} emission associated with M31) and $-52.96$ to $-44.72$ km s$^{-1}$ (Galactic foreground). For structure D, M31 emission is seen in channels $-127.16$ to $-77.70$ km s$^{-1}$, whereas in channels $-69.45$ to $-44.72$ km s$^{-1}$ the {\sc Hi} emission from structure D is connected (spatially and in velocity) to both M31 emission and large-scale Galactic emission.  

The two structures that we have labelled as {\bf A} and {\bf B} are probably mostly due to Galactic foreground. Indeed, as  Fig. \ref{fig:nnne_all_jpeg} shows, there is a tiny amount of M31 {\sc Hi} emission at the position of structure A in channels $-135.41$ and $-94.18$ to $-69.45$, but there is relatively bright Galactic {\sc Hi} emission in channels $-52.96$ to $-19.98$. Similarly, {\sc Hi} at the position of structure B is seen in channels $-52.96$ to -44.72 and $-28.23$ to $-19.98$. Because of the amount of extended Galactic {\sc Hi} emission all over the map, it is probably foreground emission but, because of its position, it cannot be excluded that M31 contributes to some of the observed IR emission from that region as well.

The unprecedented extent of our maps, together with the sensitivity of SPIRE, makes it possible to detect 3 further structures, extending to the south-west, that are most probably physically connected to M31. They appear to emanate from the outermost ring, going out to a distance of $\sim 21$, $\sim 26$ and $\sim 31$ kpc from the nucleus (see Fig. \ref{fig:herschelimages}, \ref{fig:ident_paper} and \ref{fig:EFG}, where they are named {\bf E}, {\bf F} and {\bf G}, respectively), and have been clearly detected in the {\sc Hi} map as well. The extent of these structures goes well beyond the galaxy's optical radius \citep[about 20.5 kpc, as estimated by][]{devaucouleurs91}. While the identification of ``E'' and ``F'' is quite clear at all SPIRE bands, the detection of structure ``G'' is more problematic, as it is very faint. Nonetheless, the analysis of the {\sc Hi} maps clearly shows an emission which overlaps with all three structures, at velocities that are consistent with those of Andromeda in that region, making us confident also about the presence of the outermost, faintest one. Exploiting the velocity information carried by {\sc Hi} data is in this region much simpler, as the south-west side of M31 has velocities in the opposite direction compared to those of the Galactic emission, so that contamination by foreground Galactic {\sc Hi} is not an issue. Contours defining these three structures are shown in Fig. \ref{fig:EFG}, overlaid to the high resolution {\sc Hi} map.

\begin{figure*}
\begin{tabular}{lll}
\includegraphics[width=0.317\textwidth]{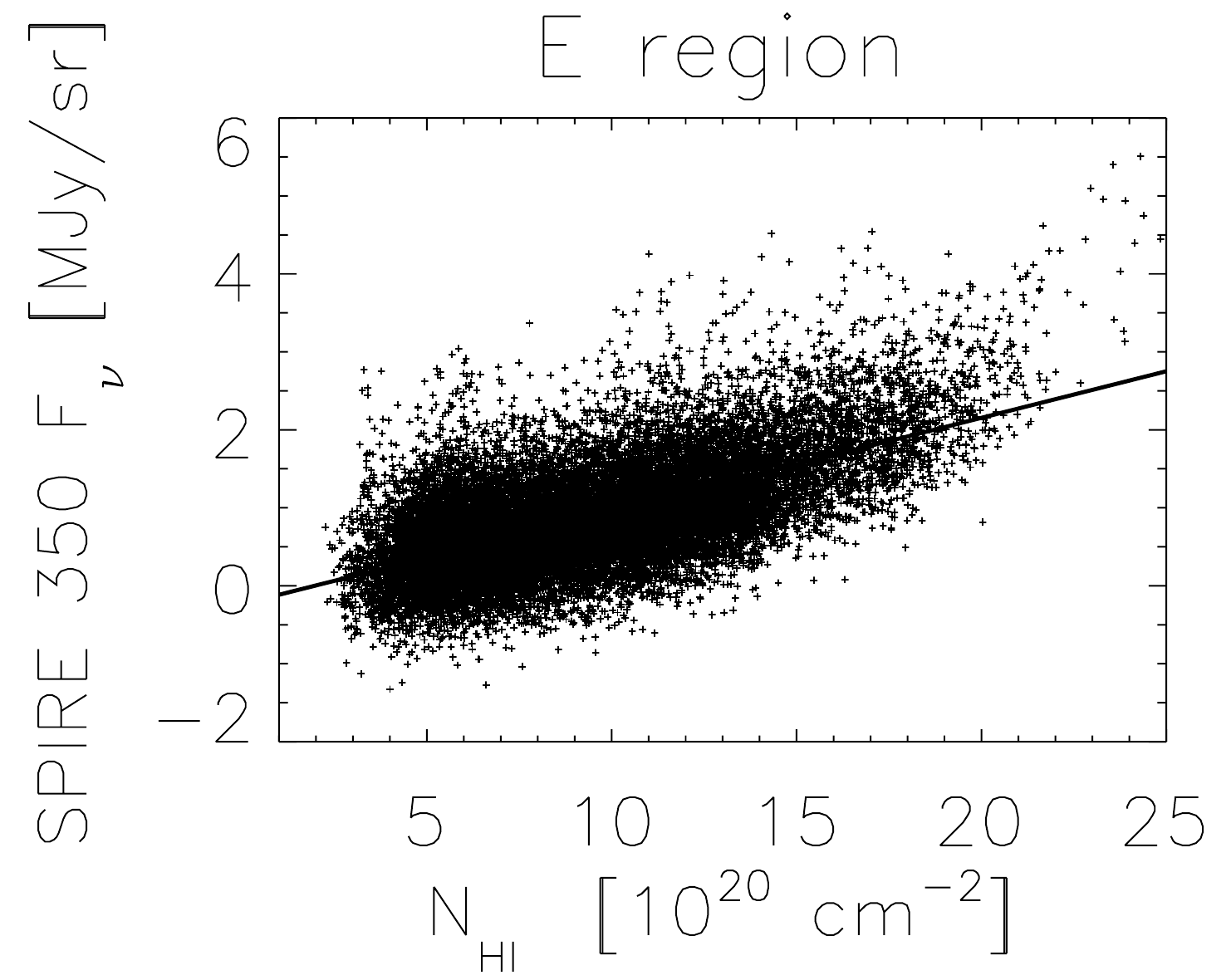} &
\includegraphics[width=0.317\textwidth]{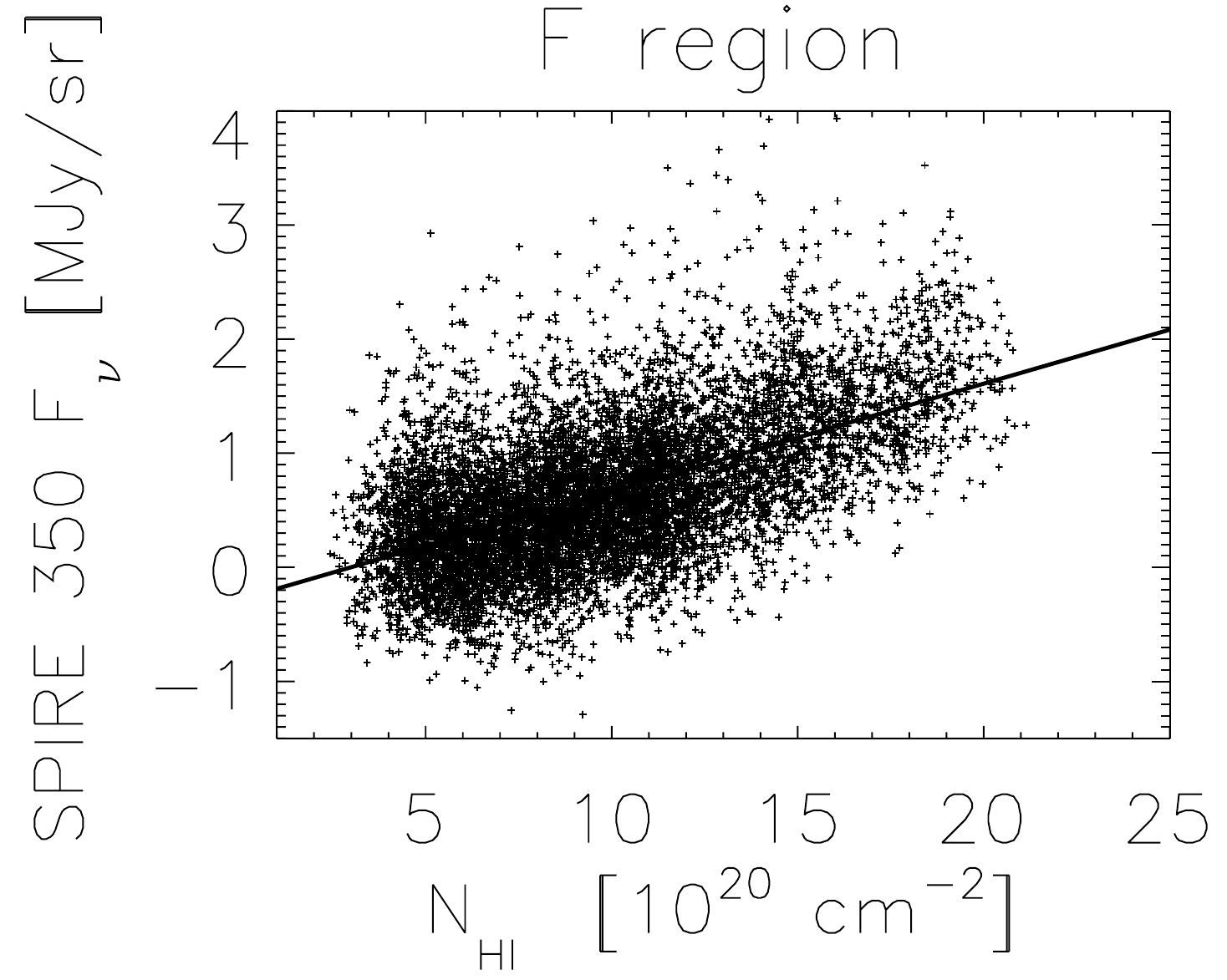} &
\includegraphics[width=0.317\textwidth]{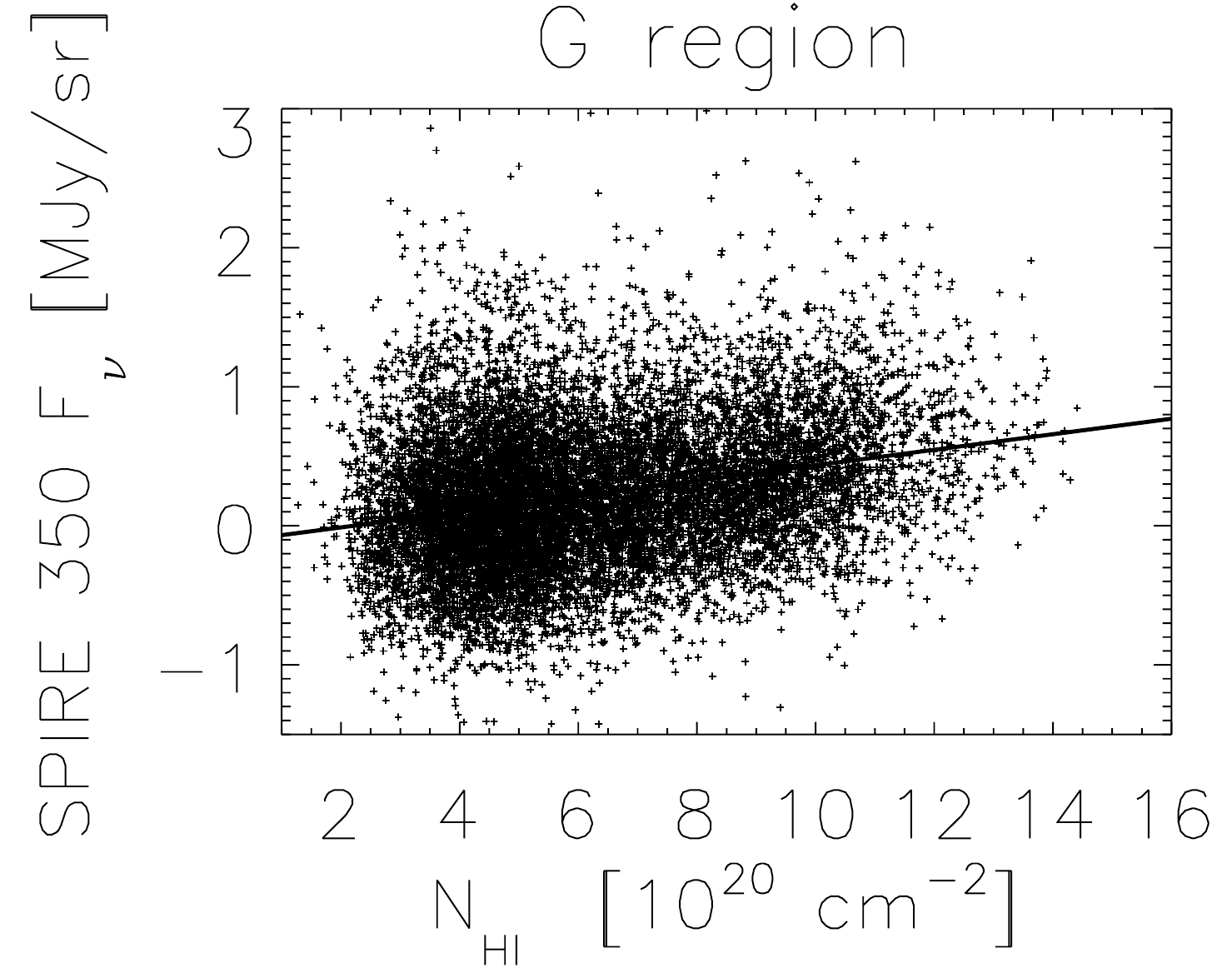} \\
\end{tabular}
\caption{Pixel-by-pixel correlation between {\sc Hi} column density, and the flux at 350 \mums for regions E, F and G (see Figs. \ref{fig:herschelimages}, \ref{fig:ident_paper} and \ref{fig:EFG}), from left to right, respectively, derived from the foreground-corrected maps. The black lines represent a linear fit to the data.}
\label{fig:EFGcorr}
\end{figure*}

\begin{figure}
\begin{center}
\includegraphics{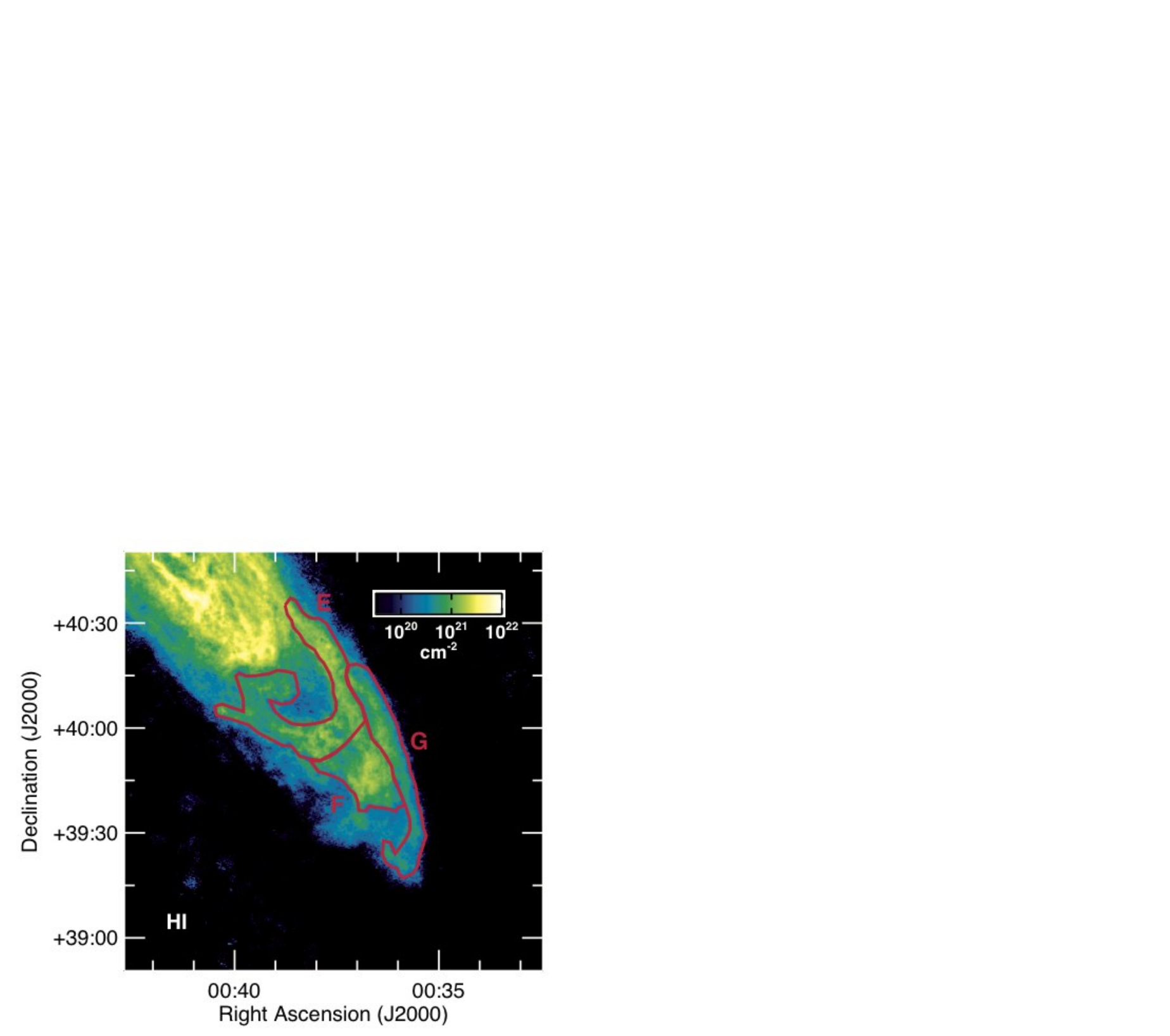}
\end{center}
\caption{A zoom on the outermost part of the south-west region in the high resolution {\sc Hi} map. Overlapped are the contours defining the three outermost regions (see text for details on the selection on these regions).}
\label{fig:EFG}
\end{figure}

In order to quantify in a more objective way the detection of these three structures we followed the approach adopted by \cite{gordon09} who, to this aim, used the correlation expected between hydrogen column density and foreground subtracted FIR maps. We have looked for such correlations in regions E, F and G, between the high resolution hydrogen map, and SPIRE images. We selected these three regions from the higher signal--to--noise {\sc Hi} map based on visually inspecting the morphology of regions with {\sc Hi} column density values greater than $\sim4\times 10^{20}$ cm$^{-2}$ (see Fig. \ref{fig:EFG}). Hence, the exact outline of these regions is, to some degree, arbitrary. We present the results for the three different regions in Fig. \ref{fig:EFGcorr}, where only SPIRE 350 \mums is shown as a representative example. Similar results are obtained when considering the other two bands, confirming our visual identification. The correlation at PACS 160 \mums is poorer and shows a much larger dispersion and is hence not used; at 100 \mums we found none.

While the whole structure of the two rings at 10 and 15 kpc is visible in all of our maps and can be easily followed all along the five {\it Herschel}'s maps, these three newly identified structures do not appear to extend all across the galaxy, possibly because of their very nature, but also due to their lower surface brightness and the high contamination from Galactic cirrus on the north-east side of Andromeda. Despite this, we note that the outermost ring--like structure on the northeast side, which is highlighted by the 250 \mums contour in Fig. \ref{fig:nnne_all_jpeg}, is located at roughly the same galacto-centric distance as structure E, on the opposite side of the galaxy.

We have also attempted to look for the possible presence of dust corresponding with the {\sc Hi} high-velocity cloud analogues likely associated with M31, as identified by \cite{thilker04}. However, after a visual inspection of the SPIRE maps, we were unable to identify any obvious signature of dust emission corresponding to any of the clouds (see Fig. \ref{fig:ident_paper}). As a further check, we have looked at the higher resolution WSRT data for the M31 High Velocity Clouds (HVCs) presented  in \cite{westmeier05}, to search for local correlations between IR and {\sc Hi} peaks.  Again, we found no obvious IR counterparts. On the one hand, we may not find any dust emission in these regions possibly because of some residual contamination from the foreground dust in some regions (especially towards the northeast). On the other hand, the foreground can be very variable, and the expected emission from these {\sc Hi} clouds is weaker than the foreground. We have indeed checked that the peak {\sc Hi} column density of Davies' cloud (see Fig. \ref{fig:ident_paper}), which is the brightest cloud found by \cite{thilker04}, is much weaker than the foreground at that position (0.4 $\times$ $10^{20}$ atoms cm$^{-2}$ and 8.1 $\times$ $10^{20}$ atoms cm$^{-2}$, respectively).

\section{Dust characteristics across the galaxy}\label{sec:dust}
\subsection{Dust masses and temperatures}\label{sec:mass_temp}
\begin{figure*}
\begin{tabular}{ll}
\includegraphics[width=0.47\textwidth]{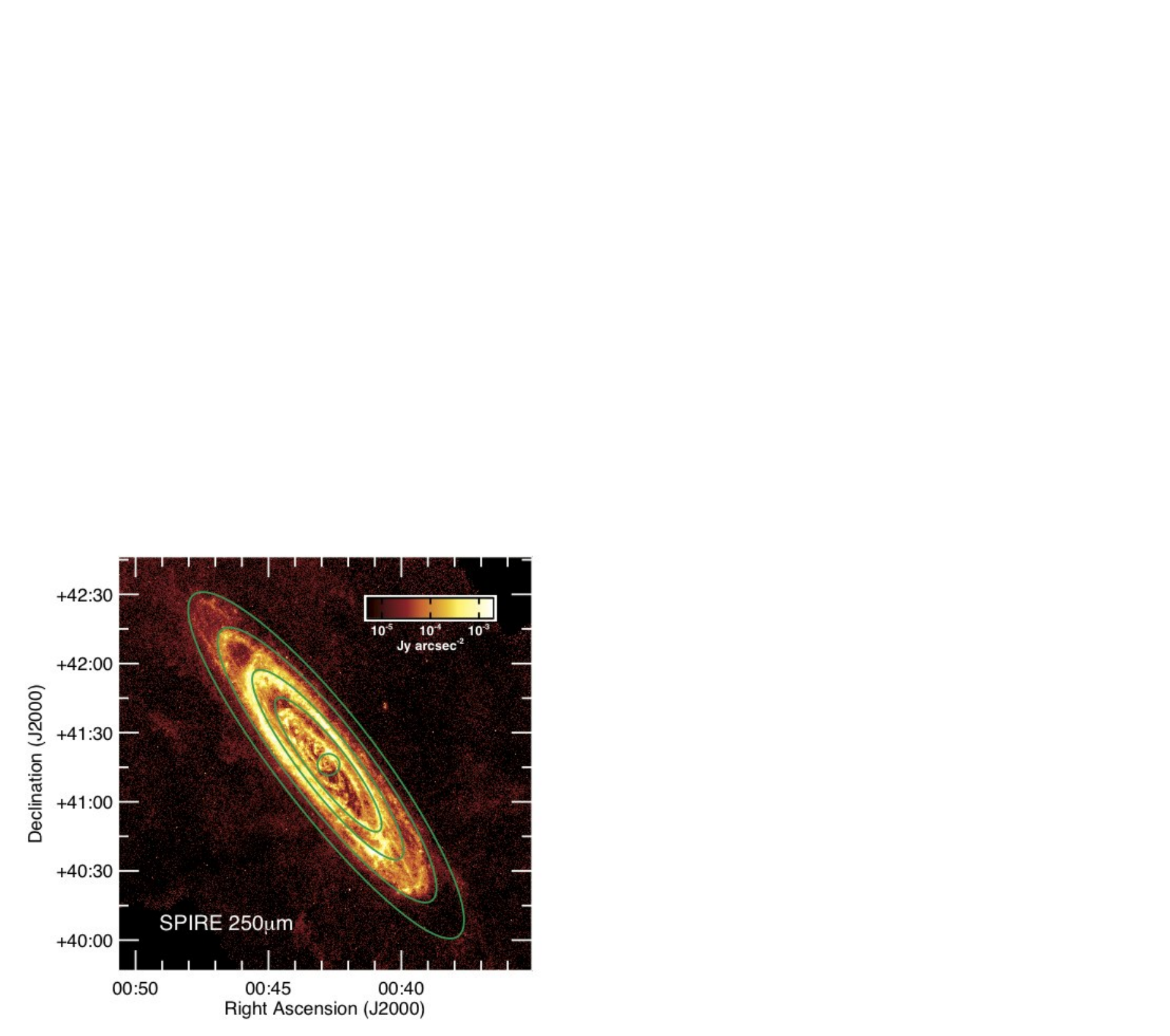} &
\includegraphics[width=0.53\textwidth]{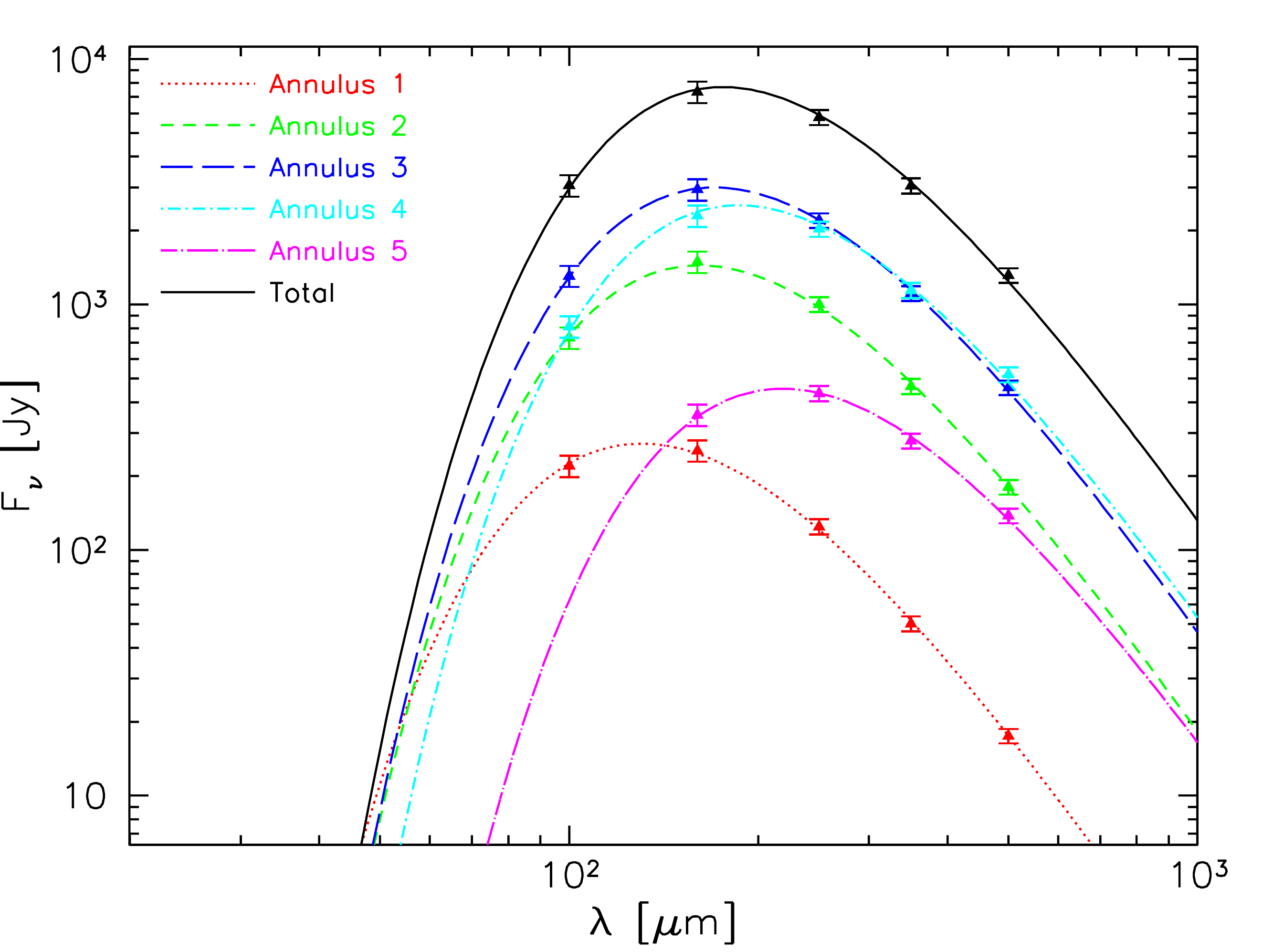}\\
\end{tabular}
\caption{In the {\it left-hand} panel the 250 \mums image is shown, together with the 5 apertures where fluxes were measured. The total aperture corresponds to the outermost elliptical aperture. The best fit values and uncertainties are given in Table \ref{tab:fitres}. In the {\it right-hand} panel we show the fit to the observed SEDs as measured in the four apertures: observed fluxes (triangles) are superimposed to the modified black body curve (lines), corresponding to the best fit values of the temperature. Different colours and lines correspond to different apertures/annuli.}
\label{fig:sedfit}
\end{figure*}

While we refer the reader to forthcoming papers \citep{smith12,ford12} performing a more detailed analysis and discussion of the dust properties, here we give a broad overview of the characteristics of dust in Andromeda, both as a function of the galacto-centric distance and considering the whole galaxy as well. This approach is similar to what has already been done by other studies performed on the global, large scale IR emission \citep[e.g.][]{walterbos87,haas98,montalto09}. To this aim, we extract fluxes in five annuli (see left-hand panel of Fig. \ref{fig:sedfit}), of which we report the sizes and characteristics in Table \ref{tab:fitres}: the first is a circular one, and it includes the innermost dust ring ($R \sim 1.5$ kpc). The sizes of the other four were obtained by projecting circular apertures that define the inner and outer boundaries of the 10 and 15 kpc rings, plus an outermost aperture including most of the structures visible in the SPIRE 250 \mums map. Their positions were chosen so that their centres coincide with the physical centres of the structures they are meant to identify \citep[see][]{block06}. As a value for the position angle, we took that of M31 ($38^\circ$). As {\it Herschel} maps differ both in resolution and pixel size, the PSFs in all maps should have in principle been matched to the broadest PSF in any map (i.e. the 500 \mum PSF). We found this to be not necessary as the apertures we have used are much larger than any of the pixel and beam sizes.

We then performed a simple SED fitting assuming that the dust thermal emission is well represented by a modified black body at a single temperature, with the emissivity being described by the following: $k_\nu = k_0\nu^\beta$, with $k_0$ such that $k=0.192$ m$^2$ kg$^{-1}$ at 350 \mums \citep{draine03}. 

We could have also used MIPS flux at 70 \mums from \cite{gordon06} but, as found by other works, this band is likely to be more sensitive to warmer dust, and would have required the addition of a further modified black body component, at higher temperature. Hence, the 70 \mums flux was only used as an upper limit. We let the temperature be a free parameter in our fitting procedure, while the value of $\beta$ was kept fixed, and initially assumed to be 2, which is consistent with the value commonly used in dust models \citep[][]{draine84,li01}. The dust mass is computed by normalizing the model SED to {\it Herschel}'s observed datapoints. Uncertainties on the fluxes were conservatively assumed to be 10\% for PACS and 7\% for SPIRE (see \S \ref{sec:pacs} and \ref{sec:spire}).
In the data fitting, the results of which we report in Table \ref{tab:fitres}, we have found the best--fitting temperature by exploiting a gradient search method. Uncertainties have been computed by bootstrapping over 200 SED fits, as follows: once the best fit solution is found a set of new, fictitious, datapoints is created by randomly choosing a value within the observed errorbars. A best fit model is found for these new flux values, and the procedure is repeated 200 times. The 16\% of the best fit models with the lower and higher values are discarded, and uncertainties are then taken as the differences between the best fit solution and the extreme values.

\begin{table*}
\centering
\begin{tabular}{lrrrrrrrrrrrrrr}
\hline
\hline
Ap          &     100     &     160     &   250      &    350   &     500   &    {\sc a}     &  {\sc b}    &           T$_D$             &              M$_D$                &    M$_{\rm HI}$       &    M$_{\rm HI}^*$       &   M$_{\rm H_2}$    & {\sc g2d} &     {\sc g2d}$^*$      \\
\hline 
\T   
               &    Jy     &    Jy      &  Jy     &     Jy    &   Jy    &   kpc    &    kpc    &               K                &          lg(M$_\odot$)          & lg(M$_\odot$)  & lg(M$_\odot$)  & lg(M$_\odot$) &            &            \B\\
\hline  
\T 
1             &   220  &      254  &    124 &      50 &       18  &    2.2     &    2.2    & $22.2^{+0.4}_{-0.5}$ & $ 5.66^{+0.03}_{-0.03}$ &    7.19    &    7.21    &       6.82     &     66              &      68        \B \\
2             &   732  &    1493  &  1001 &   464  &    181  &   16.4    &    3.6    & $18.0^{+0.3}_{-0.3}$ & $ 6.85^{+0.03}_{-0.03}$ &    8.48    &    8.54    &      7.93     &      75             &      83        \B \\
3             & 1307  &    2943 &  2200 & 1110  &    459  &   23.2    &    5.2    & $17.0^{+0.3}_{-0.3}$ & $ 7.29^{+0.03}_{-0.03}$ &    9.03    &    9.13    &      8.40      &      92             &    111        \B \\
4             &   813  &    2302 &  2031 & 1143  &    518  &   33.6    &    7.5    & $15.6^{+0.2}_{-0.3}$ & $ 7.41^{+0.04}_{-0.03}$ &    9.31    &    9.42    &        ---        &    $^a$109    & $^a$140  \B \\
5             &   ---     &      354  &    435 &    278 &     138 &   42.4    &    9.5    & $13.3^{+0.4}_{-0.3}$ & $ 7.01^{+0.05}_{-0.04}$ &    8.99    &    9.08    &        ---        &    $^b$131    & $^b$158  \B \\
\sc{tot}   & 3055  &    7348 &  5791 &  3047 &  1313  &   42.4    &    9.5    & $16.5^{+0.2}_{-0.3}$ & $ 7.76^{+0.04}_{-0.03}$ &    9.65   &    9.74     &      8.42      &     111            &     137       \B \\
\hline
\hline
\B
\end{tabular}
\caption{{\it Herschel} fluxes, as measured within the five annuli we have considered (see Fig. \ref{fig:sedfit}, left-hand panel). Dust temperature (T$_D$) and masses (M$_D$), and their uncertainties, are derived from SED fitting (see text for details).  We also report the mass of hydrogen in the atomic phase, as derived from {\sc Hi} maps of \cite{braun09} (both corrected and non--corrected for self--opacity), and the mass of gas in the molecular phase derived by CO data \citep{nieten06}. Last two columns give the total gas--to--dust ratios within the given annulus. Results for the largest annulus and for the whole galaxy are limited by the size of the CO map (see text for more details): the value indicated with $^a$ assumes a mass of 5\% of total, within that annulus, and the value of the gas--to--dust in the outermost annulus ($^b$) was computed assuming a negligible contribution from the molecular fraction, and it is hence to be interpreted as a lower limit. Values in the columns labeled with $^*$ are calculated using the self--opacity corrected maps. ``{\sc a}'' and ``{\sc b}'' are the major and minor axis length, respectively, of the ellipses, given in kpc (first aperture is circular).}
\label{tab:fitres}
\end{table*}

We compare the value of the dust mass we have obtained to the one derived by \cite{haas98}, who use a similar approach as ours in their analysis of ISO data. If we consider the dust included within the fourth annulus, which is the one that better matches their global aperture, we obtain a total mass of $\sim 5.3\times 10^7$ M$_\odot$ which is almost a factor of two higher than their $3\times 10^7$ M$_\odot$ value, and also higher than the value derived by \cite{gordon06} ($4\times 10^7$ M$_\odot$, at a temperature of 17 K). The temperature they obtain from the global SED --16 K-- agrees fairly well with the value of 16.8 K that we would derive fitting the total flux within the fourth annulus, and is somewhat lower than the 18.7 K value found by \cite{tabatabaei10}. The difference in the dust mass value might be, at least partially, due to the use of different dust emissivity coefficients \citep[Haas et al. use the][prescription for calculating the dust mass]{hildebrand83}. Furthermore, the size scales used in determining the total dust mass also affect the uncertainty in the total dust mass as shown by \citet{galliano11} to be up to $\sim50$\%, for example, in the Large Magellanic Cloud. The value of this uncertainty will depend on the morphology of the galaxy and the mixing of the different dust phases in the area used for total dust mass determination. 

We have also explored different values of $\beta$, namely 2.5, 1.7 and 1.5. $\beta=2$ turned out to be the best suited value for reproducing the annuli SEDs. Only if we consider the total aperture, fits with an acceptable $\chi^2$ were obtained also for $\beta=1.7$, yielding in this case a temperature of 17.7 K and a total dust mass of $4.7\times 10^7$ M$_\odot$.

The variation of the average temperature --as a function of the radius-- derived with this analysis is similar, within the errorbars, to that found by \cite{tabatabaei10}, derived from MIPS 70 and 160 \mums images, with the exception of the central regions in which the temperature we found is lower, most likely due both to our coarser sampling, and to the fact that we are not fitting the MIPS 70 \mums point, which would be more sensitive to warmer dust.

As already found in previous studies \citep[e.g.][]{haas98,gordon06}, the dust emission in the 1.5 kpc ring region is dominated by a higher temperature component. This is also consistent with the study by \cite{devereux94}, who find that the gas in the innermost regions is dominated by an ionized component (seen in H$\alpha$ emission). On the other hand, if we only consider the emission from the 10 kpc ring (sampled by our third annulus), we obtain an average temperature of $17.0^{+0.3}_{-0.3}$ K, and a dust mass of $1.95\times 10^7$ M$_\odot$. Our global SED fitting shows no need for an extra, very cold, dust component that, revealing its presence only at SPIRE wavelengths, would have been impossible to detect by former studies exploiting only data up to $\sim170$ \mum. This issue is tackled in a better detail in a companion paper \citep{smith12}, performing a pixel--by--pixel analysis.

\subsection{The gas--to--dust ratio}
We have calculated the mass of neutral hydrogen from the {\sc Hi} maps, within the same annuli, and present it in Table \ref{tab:fitres}. To take the atomic Helium fraction into account, this value is multiplied by a factor of 1.36.

The CO map was used to derive the mass of molecular hydrogen by assuming a H$_2$-CO conversion factor  X = N(H$_2$)/I$_{CO} =1.9 \times 10^{20}$ cm$^{-2}$ (K km s$^{-1}$)$^{-1}$ \citep{strong96}. The CO--to--molecular hydrogen conversion factor is notoriously affected by a large intrinsic uncertainty. Values derived for M31, in particular, can range from $0.97\times 10^{20}$ to $4.6\times 10^{20}$ cm$^{-2}$ (K km s$^{-1}$)$^{-1}$ \citep{leroy11}. In \cite{smith12} we have performed a detailed, spatially resolved analysis of dust emission considering all {\it Herschel} bands, and we have found values, for the X$_{CO}$ factor in the range $1.5\div 2.3 \times 10^{20}$, with a best-fit value of $1.9\times 10^{20}$ cm$^{-2}$ (K km s$^{-1}$)$^{-1}$. 

The total molecular gas mass ($M_{\rm mol}=M_{H_2}+M_{He}$) is then calculated by multiplying the H$_2$ mass value by 1.36. The CO observations only sample the region which includes the 10 kpc ring. \cite{nieten06} estimate that, within a region covering the innermost $\sim 18$ kpc, approximatively corresponding to our fourth annulus, a further 5\% of CO mass is expected. In calculating the gas--to--dust ratio in the two outermost apertures (and of course in the total one), we have applied this correction, also assuming that no further molecular gas is expected at larger galacto-centric distances. Hence, the values of the gas--to--dust ratio for the aperture 4, 5 and the total, that we give in Table \ref{tab:fitres}, and that already include this correction, should be considered as lower limits, although not too distant from the actual values. In fact, it should be noted that the gas mass is dominated by the atomic phase, the molecular only contributing to the total by less than $\sim10$\%.

That the gas--to--dust ratio in M31 is not constant throughout the galaxy, and increases as a function of the radial coordinate, was already shown by several previous studies \citep[see e.g.][]{bajaja77,walterbos88,nedialkov00}. We estimated an average value of $\sim 110$, calculated over the largest aperture ({\sc g2d} column of Table \ref{tab:fitres}), and a variation by a factor of $\sim 2$ from the innermost ring to the outermost annulus.

We found a very good correlation between the gas--to--dust ratio and the (average) galacto-centric distance, with the first quantity exponentially increasing from the centre to the outskirts. Even though our analysis is performed over very broad apertures, we found a gradient of 0.0163 dex/kpc, which is consistent with the values of the metallicity gradient found by \cite{trundle02} in M31, ranging from -0.027 to -0.013 dex/kpc, under the assumption that the amount of metals locked up into dust grains is a constant fraction of the mass of interstellar metals \citep{edmunds01}. Should this be the case, then the metallicity gradient would be the opposite as the one for the gas--to--dust.

\cite{braun09} also provide maps where hydrogen self--opacity is taken into account. It is quite clear that the simple 21 cm integral is a  lower limit to the {\sc Hi} mass, and M31 is one of the few cases where we have the data to account for this problem \citep{braun12}. In fact, had we taken this into account, the atomic mass would have increased by a factor ranging from $\sim 5$\% in the innermost regions, to $\sim 27$\% in the most distant annulus. In Table \ref{tab:fitres} we report the {\sc Hi} mass and the gas--to--dust ratio calculated from the self--opacity corrected maps (columns indicated with a $^*$).

\subsection{Dust in the outskirts}\label{sec:outskirts}
We will now briefly characterize the properties of dust that is detected in the outermost regions. In Sec. \ref{sec:identification} and \ref{sec:fir_struct} we have identified structures in the outskirts of M31 disk, that are likely to belong to Andromeda. As even after subtracting a model for the Galactic cirrus emission the situation remains confused in the North--East side, we will focus on the structures that are identified on the southern part, as they are much less affected by foreground dust emission and which we have labelled as E, F and G.

In Sec. \ref{sec:fir_struct} we have shown that there is a significant correlation between the hydrogen column density and FIR emission at SPIRE wavelengths for these arc--like shaped structures. Using the same technique that is described in Sec. \ref{sec:dust}, we fit the observed fluxes measured within the areas defined by these regions (see Fig. \ref{fig:EFG}), by means of a modified black body SED. As there is no clear detection at PACS wavelengths, we are forced to limit ourselves to SPIRE wavelengths. 

As before, we have considered four values for the emissivity spectral index $\beta$, namely 1.5, 1.7, 2.0 and 2.5. Acceptable fits were achieved with all the four values of $\beta$ for region E, with $\beta=1.5$ being slightly preferred. The SEDs of regions F and G turned out to be compatible with $\beta=1.5$ and 1.7 only, with the first value yielding, again, slightly better fits. Considering only $\beta=1.5$, we found dust masses in these three regions of $1.23\times 10^6$ M$_\odot$, $5.89\times 10^5$ M$_\odot$ and $3.55\times 10^5$ M$_\odot$, respectively, and temperatures of 14.4, 13.1 and 11.6 K. Measuring the hydrogen mass on the same regions, we derive {\sc Hi}--to--dust ratios of 202, 204 and 274, respectively. 

\section{Summary and conclusions}\label{sec:conclusions}
We have presented and described {\it Herschel} PACS (100 and 160 \mum) and SPIRE (250, 350 and 500 \mum) observations of a $\sim 5.5^\circ \times 2.5^\circ$ field centered on the local galaxy M31. Using high spatial and velocity resolution {\sc Hi} maps, roughly covering the same area, we have identified the dusty structures that surround the galaxy outside the already well-observed $\sim 3^\circ \times 1^\circ$ region \citep{haas98,gordon06,barmby06,krause11}. 

We have performed a broad analysis of the characteristics of the dust as a function of the radius. We have considered 5 main regions, described by projected circular apertures (plus an elliptical one), sampling the most significant morphological structures: the inner 1.5 kpc ring, the intra-arm region enclosed within the 10--kpc ring, the 10--kpc ring itself, the outer ($\sim15$ kpc) ring, and a fifth one, containing the outermost regions. A simple one--temperature modified black body model is able to successfully reproduce the observed PACS/SPIRE SED. Average quantities, such as the dust temperature, and the gas--to--dust ratio, are derived, and they are found to strongly depend on the radius. The gas--to--dust ratio, in particular, is changing by a factor of $\sim 2$, following an exponential increase, from the inner regions to the outskirts, as expected from the presence of a metallicity gradient with the opposite sign, across the galaxy. 

According to our simple emission model the two main rings, at 10 and 15 kpc from the centre, contain the bulk (about 78\%) of the dust in all the galaxy. The dust surface density is the highest in within the 10 kpc one, 0.41 M$_\odot$/pc$^2$, being higher by a factor of $\sim 4$ with respect to the average value, and also higher than in the 15 kpc ring, where it is about $0.25$ M$_\odot$/pc$^2$. 

The dust temperature is highest in the innermost aperture, but the value we derive from our SED fitting, 22 K, is considerably lower than previous studies  \citep[e.g.][derive temperatures of $\sim 28$ and $\sim 30$ K, respectively, for the nuclear dust]{haas98,tabatabaei10}. This is most likely due to the fact that, ignoring data shortwards 100 \mum, we are not sensitive to the presence of a warmer component \citep[e.g.][]{bendo10}: this would, anyway, constitute a minor fraction of the total dust content.

We will give a more detailed description of the IR morphology in a forthcoming paper \citep{kirk12}, where an analysis of the structures on deprojected {\it Herschel} maps will be performed.

The maps we describe in this work present two distinct situations: in the north--eastern part, the combination of Andromeda's systemic and rotation velocities, yielding values that are similar to those expected for the interstellar medium in our galaxy, together with the presence of a substantial amount of gas and dust, makes the identification of emission regions that actually belong to Andromeda very difficult, and sometimes ambiguous. On the other hand, at the southern side, not only is the velocity of gas and dust that belong to Andromeda expected to have a clearly different pattern with respect to that of our Galaxy, but is also much less affected by the presence of interstellar medium (see e.g. Fig. \ref{fig:ident_paper} for a panoramic view of the M31 neighbourhood). 

In this region, we identify in our maps three different ring-shaped, concentric structures -- which we label E, F and G --  that can be safely attributed to M31, based on the shape and velocity of their {\sc Hi} counterparts, extending out to a radius of $\sim 31$ kpc. Dust in the outskirt of Andromeda had been already indirectly detected, through extinction of both stars and background galaxies, by \cite{cuillandre01}, who exploited deep optical photometry in a $\sim 28' \times 28'$ field roughly centred on our ``E'' region. This result, recently confirmed on a smaller scale by \cite{bernard12} (their ``Outer Disc field''), implies an atomic gas--to--dust ratio about three times higher with respect to the average Galactic value, and is considered as an upper limit since it only takes into account the dust in front of the observed stars. Instead, the {\sc Hi}--to--dust ratio we estimate for this particular region, $\sim 204$, is a factor of $\sim 1.5$ larger than the 137 value estimated for the MW  \citep[see Table 2 in][]{draine07}, that one would expect given the roughly solar metallicity of stars observed in a nearby field. This value can provide a more stringent constraint to the amount of dust in this region, and to the chemical enrichment of the interstellar medium in Andromeda's outskirts.

\section*{Acknowledgements}

We would like to thank the anonymous referee, as her/his comments led to a substantial improvement of our first manuscript.\\
MB, JF, IDL and JV acknowledge the support of the Flemish Fund for Scientific Research (FWO-Vlaanderen), in the frame of the research projects no. G.0130.08N and no. G.0787.10N.\\
GG is a postdoctoral researcher of the FWO-Vlaanderen (Belgium). \\
The research leading to these results has received funding from the European Community's Seventh Framework Programme (/FP7/2007-2013/) under grant agreement No 229517.\\
The research of C.D.W. is supported by grants from the Natural Sciences and Engineering Research Council of Canada.\\
SPIRE has been developed by a consortium of institutes led by Cardiff University (UK) and including Univ. Lethbridge (Canada); NAOC (China); CEA, LAM (France); IFSI, Univ. Padua (Italy); IAC (Spain); Stockholm Observatory (Sweden); STFC and UKSA (UK); Imperial College London, RAL, UCL-MSSL, UKATC, Univ. Sussex (UK); and Caltech, JPL, NHSC, Univ. Colorado (USA). This development has been supported by national funding agencies: CSA (Canada); NAOC (China); CEA, CNES, CNRS (France); ASI (Italy); MCINN (Spain); Stockholm Observatory (Sweden); STFC (UK); and NASA (USA).\\
PACS has been developed by a consortium of institutes led by MPE (Germany) and including UVIE (Austria); KU Leuven, CSL, IMEC (Belgium); CEA, LAM (France); MPIA (Germany); INAFIFSI/OAA/OAP/OAT, LENS, SISSA (Italy); IAC (Spain). This development has been supported by the funding agencies BMVIT (Austria), ESA-PRODEX (Belgium), CEA/CNES (France), DLR (Germany), ASI/INAF (Italy), and CICYT/MCYT (Spain).
%%%%%%%%%%%%%%%%%%%%%%%%%%%%%%%%%%%%%%%%%%%%%%%%%%%%%%%%%%%%%%%%%%%%%%%%%%%%%%%%%%%%

\end{document}